\RequirePackage{etoolbox}
\csdef{input@path}{%
 {sty/}
 {img/}
}%
\csgdef{bibdir}{bib/}

\documentclass[ba]{imsart}
\pubyear{0000}
\volume{00}
\issue{0}
\doi{0000}
\firstpage{1}
\lastpage{1}

\usepackage{amsthm, amsmath, amssymb,xspace,graphics}
\usepackage{natbib}
\usepackage[colorlinks,citecolor=blue,urlcolor=blue,filecolor=blue,backref=page]{hyperref}
\usepackage{graphicx}
\usepackage{algpseudocode}
\usepackage{algorithm}
\usepackage{media9}
\usepackage{subfig}
\usepackage{booktabs}
\newcommand{\ra}[1]{\renewcommand{\arraystretch}{#1}}

\newcommand{\D}{\text{D}}
\newcommand{\pro}{_{\text{prop}}}
\newcommand{\obs}{_{\text{obs}}}
\newcommand{\dist}{_{\text{dist}}}
\newcommand{\init}{_{\text{init}}}

\startlocaldefs
\numberwithin{equation}{section}
\theoremstyle{plain}

\endlocaldefs

\begin{document}

\begin{frontmatter}
\title{Adaptive Approximate Bayesian Computation Tolerance Selection}
\runtitle{Adaptive Approximate Bayesian Computation Tolerance Selection}

\begin{aug}
\author{\fnms{Umberto} \snm{Simola}\thanksref{addr1,t1,t2,m1}\ead[label=e1]{umberto.simola@helsinki.fi}}
\and
\author{\fnms{Jessi} \snm{Cisewski-Kehe}\thanksref{addr2,t3,m1,m2}\ead[label=e2]{jessica.cisewski@yale.edu}}
\and
\author{\fnms{Michael U.} \snm{Gutmann}\thanksref{addr3,t3,m1,m2}\ead[label=e3]{michael.gutmann@ed.ac.uk}}
\and
\author{\fnms{Jukka} \snm{Corander}\thanksref{addr1,addr4,t3,m1,m2}\ead[label=e4]{jukka.corander@medesin.uio.fi}}

\runauthor{U. Simola et al.}

\address[addr1]{ Department of Mathematics and Statistics, University of Helsinki, Helsinki, Finland
    \printead{e1} 
}
\address[addr2]{Department of Statistics and Data Science, Yale University, New Haven, CT, USA
    \printead*{e2}
}
\address[addr3]{School of Informatics, University of Edinburgh, Edinburgh, United Kingdom
    \printead*{e3}
}
\address[addr4]{Department of Biostatistics, University of Oslo, Oslo, Norway
    \printead*{e4}
}


\end{aug}

\begin{abstract}
Approximate Bayesian Computation (ABC) methods are increasingly used for inference in situations in which the likelihood function is either computationally costly or intractable to evaluate.
Extensions of the basic ABC rejection algorithm have improved the computational efficiency of the procedure and broadened its applicability.  The ABC - Population Monte Carlo (ABC-PMC) approach has become a popular choice for approximate sampling from the posterior.  ABC-PMC is a sequential sampler with an iteratively decreasing value of the  tolerance, which specifies how close the simulated data need to be to the real data for acceptance.  We propose a method for adaptively selecting a sequence of tolerances that improves the computational efficiency of the algorithm over other common techniques. In addition we define a stopping rule as a by-product of the adaptation procedure, which assists in automating termination of sampling. The proposed automatic ABC-PMC algorithm can be easily implemented and we present several examples demonstrating its benefits in terms of computational efficiency.
\end{abstract}

\begin{keyword}
\kwd{Complex stochastic modeling, likelihood-free methods, sequential Monte Carlo}
\end{keyword}

\end{frontmatter}

\section{Introduction} \label{sec:intro}
\begin{sloppypar}
Approximate Bayesian Computation (ABC) provides a framework for inference in situations where the relationship between the data and the parameters does not lead to a tractable likelihood function, but where forward simulation of the data-generating process is possible. ABC has been used in many areas of science such as biology \citep{ThorntonEtAl2006}, epidemiology \citep{McKinleyEtAl2009, numminen2013estimating}, ecology \citep{Beaumont2010}, population modeling \citep{TonyEtAl}, modeling the population effects of a vaccine \citep{corander2017frequency}, dark matter direct detection \citep{simola2019machine}, and astronomy \citep{CameronPettitt2012, Cisewski-Kehe:2019aa, IshidaEtAl2015, ShaferFreeman2012, WeyantEtAl2013}. 
The basic ABC algorithm \citep{PitchardEtAl1999,Rubin1984,TavareEtAl1997} can be explained in four steps. 
Suppose the parameter vector $\theta \in \mathbb R^p$ is the target of inference, then 
(i) draw the model parameters from the prior distribution, $\theta\pro \sim \pi(\theta)$, 
(ii) produce a synthetic sample of the data by using $\theta\pro$ in the forward simulation model, $y\pro \sim f(y \mid \theta\pro)$, 
(iii) compare the true data, $y\obs$, with the generated sample, $y\pro$, using a distance function, $\rho(\cdot, \cdot)$, and defining the distance as $d = \rho(s(y\obs), s(y\pro))$ where $s(\cdot)$ is some (possibly multi-dimensional) summary statistic of the data, 
(iv) if the distance, $d$, is less than or equal to a fixed tolerance, $\epsilon$, then $\theta\pro$ is retained, otherwise it is discarded. This is repeated until a desired particle sample size, $N$, is achieved. 
\end{sloppypar}

Following the notation of \cite{MarinEtAl2012}, the resulting ABC posterior can be written as 
\begin{equation*}
\pi_{\epsilon}(\theta \mid y\obs) = \int \left[ \frac{f(y\pro \mid \theta)\pi(\theta)\mathbb I_{A_{\epsilon,y\obs}}(y\pro)}{\int_{A_{\epsilon,y\obs} \times \Theta} f(y\pro \mid \theta)\pi(\theta) dy\pro d\theta}\right]dy\pro, \label{ABCposterior}
\end{equation*}
where $\mathbb I_{A_{\epsilon,y\obs}}(\cdot)$ is the indicator function for the set $A_{\epsilon,y\obs} = \{y\pro \mid \rho(s(y\obs),s(y\pro)) \leq \epsilon\}$.  
There are many extentions to the basic ABC algorithm (e.g., \citealt{Blum2010,BlumEtAl2013,RatmannEtAl2013,CsilleryEtAl2010, DelMoralEtAl2012, DrovandiEtAl2011, FearnheadPrangle2012, JoyceMarjoram2008, MarinEtAl2012}), but here we focus on the ABC - Population Monte Carlo (ABC-PMC) approach introduced by \citet{BeaumontEtAl2009}. However, the proposed methodology could be used in other sequential versions of ABC that require selecting a sequence of tolerances. The proposed adaptive approximate Bayesian computation tolerance selection algorithm (aABC-PMC) targets the same kind of approximate posterior sampling problems as the original ABC-PMC algorithm, and may be subject to the same limitations in the case of high-dimensional parameter spaces. ABC has been successfully used in numerous situations where the likelihood function is intractable and the number of parameters varies from 2 to 5 (e.g. \citealt{BeaumontEtAl2009, Cisewski-Kehe:2019aa, CsilleryEtAl2010, Cornuet2008, DelMoralEtAl2012, gutmann2016bayesian, jarvenpaa2016gaussian, JenningsEtAl, JenningsEtAl2, numminen2013estimating, Silk2013, simola2019machine, SissonEtAl2007, TonyEtAl}). Our algorithm is designed to significantly improve upon the original ABC-PMC method under similar circumstances.

The ABC-PMC algorithm by \citet{BeaumontEtAl2009} is based on an adaptive importance sampling approach, where, given a series of decreasing tolerances $\epsilon_1 > \epsilon_2 > \cdots > \epsilon_T$ (T being the final iteration), the proposal distribution is sequentially updated in order to improve the efficiency of the algorithm. This is done by constructing a series of intermediate proposal distributions, with the details of the steps presented in Algorithm~\ref{alg2:ABC-PMC}. The first iteration of the ABC-PMC algorithm uses tolerance $\epsilon_1$ and draws proposals from the specified prior distribution(s); the corresponding ABC posterior is denoted by $\pi_{\epsilon_1}$.   
Rather than starting the rejection sampling over using a smaller $\epsilon$, the algorithm proceeds sequentially by drawing proposals from the ABC posterior approximated in the previous iteration. 
After a parameter value, typically referred to as a \textit{particle}, is selected from the set of available particles from the previous iteration, it is also translocated according to some kernel function (e.g. a Gaussian kernel) to avoid degeneracy of the sampler.  Since the proposals are not drawn directly from the prior $\pi$, importance weights are used.  The importance weight for a particle $J = 1, \ldots, N$ at iteration $t$ is:
\begin{equation}
W_t^{(J)} \propto \pi(\theta_{t}^{(J)})/\sum_{K = 1}^N W_{t-1}^{(K)} \phi\left[\tau_{t-1}^{-1}\left(\theta_{t}^{(J)} - \theta_{t-1}^{(K)}\right)\right], \label{eq:importance_weights}
\end{equation}
where $\phi(\cdot)$ is the density function of a standard normal distribution\footnote{The probability density function of a $Q$--dimensional standard normal distribution is $\phi(X)= {(2\pi)}^{-\frac{Q}{2}} \exp{\left(-\frac{1}{2} X^TX\right)}$ with the expected value of the random vector X is $E[X]=\vec{0}$ (where $\vec{0}$ is a Q-dimensional vector of zeros) and its covariance matrix is $\text{Var}[X]=I_Q$, where $I_Q$ is the $Q \times Q$ identity matrix.}, $\tau_{t-1}^2$ is the variance  (twice the weighted sample variance of the particles from iteration $t-1$ is used, as recommended in \citealt{BeaumontEtAl2009}), and $\pi(\cdot)$ is the prior distribution. We note that the definition for the importance weight provided in Eq.\eqref{eq:importance_weights} is up to a normalization constant. In fact each importance weight is normalized such that $\sum_{J=1}^{N} W_t^{(J)} =1$.
While the particles are drawn from a sequentially improving proposal distribution, the tolerances also decrease such that $\epsilon_1 > \epsilon_2 > \cdots > \epsilon_T$, to increase the fidelity of the resulting approximation to the underlying posterior. The common strategies for selecting this sequence adaptively, highlighted in Section~\ref{sec:toleranceAndstop}, can lead to inefficient sampling as well as avoiding relevant regions of the parameter space \citep{Silk2013}. The key contributions of this article are (i) a method for selecting the $\epsilon_{1:T} = (\epsilon_1, \epsilon_2, \ldots, \epsilon_T)$ in a manner that results in improved computational efficiency, and (ii) a rule for determining when the algorithm terminates (i.e. determining $T$).

\begin{algorithm}
\caption{ABC-PMC algorithm for $\theta$}
\begin{algorithmic}
\State Given a series of decreasing tolerances $\epsilon_1 > \epsilon_2 > \cdots > \epsilon_T$
\If{$t = 1$}
\For{$J = 1, \ldots, N$}
\State Set $d_1^{(J)} = \epsilon_1 + 1$
 \While{$d_1^{(J)} > \epsilon_1$}
	\State Propose $\theta^{(J)}$ by drawing $\theta\pro \sim \pi(\theta)$,
 	\State Generate  $y\pro \sim f\left(y\mid \theta^{(J)}\right)$
 	\State Calculate distance $d_1^{(J)} = \rho(s(y\obs), s(y\pro))$
 \EndWhile
 	\State Set weight $W_1^{(J)} = N^{-1}$
 \EndFor
 \ElsIf{$2 \leq t \leq T$}
  \State Set $\tau_{t}^{2} = 2 \cdot \text{var} \left( \{\theta_{t-1}^{(J)},W_{t-1}^{(J)}\}_{J=1}^{N}\right)$
 \For{$J = 1, \ldots, N$}
 \State Set $d_t^{(J)} = \epsilon_t + 1$
 \While{$d_t^{(J)}> \epsilon_t$}
	\State Select $\theta_t^*$ from $\theta_{t-1}^{(J)}$ with probabilities $\left\{W_{t-1}^{(J)}/\sum_{K = 1}^NW_{t-1}^{(K)}\right\}_{J = 1}^N$
	\State Propose $\theta_t^{(J)} \sim \mathcal{N}(\theta_t^*, \tau_{t}^{2})$
	\State Generate  $y\pro \sim f\left(y \mid \theta_{t}^{(J)}\right)$
 	\State Calculate distance $d_t^{(J)} = \rho(s(y\obs), s(y\pro))$
 \EndWhile
 	\State Set weight $W_t^{(J)} \propto \pi(\theta_t^{(J)})/\sum_{K = 1}^N W_{t-1}^{(K)} \phi\left[\tau_{t-1}^{-1}\left(\theta_{t}^{(J)} - \theta_{t-1}^{(K)}\right)\right]$
 \EndFor
 \EndIf
\end{algorithmic} \label{alg2:ABC-PMC}
\end{algorithm}

\subsection{Selecting the tolerance sequence and stopping rules} \label{sec:toleranceAndstop}

There are three common approaches for selecting the tolerance sequence, $\epsilon_{1:T}$:  (i)  fixing the values in advance \citep{BeaumontEtAl2009, McKinleyEtAl2009,SissonEtAl2007, TonyEtAl}, (ii) adaptively selecting $\epsilon_t$ based on some quantile of $\{d_{t-1}^{(J)}\}_{J = 1}^N$, the distances of the accepted particles from iteration $t-1$ \citep{Cisewski-Kehe:2019aa,IshidaEtAl2015,Lenormand2013,simola2019machine, WeyantEtAl2013}, or (iii) adaptively selecting $\epsilon_t$ based on some quantile of the effective sample size (ESS) values \citep{DelMoralEtAl2012, numminen2013estimating}. These approaches can lead to inefficient sampling as discussed below and demonstrated in the simulation study in Section~\ref{sec:examples}. 
It turns out that selecting tolerances using a predetermined quantile can, if not selected wisely, lead to the particle system getting stuck in local modes \citep{Silk2013}.
Hence the exact sequence of tolerances has an impact not only on the computational efficiency of the algorithm but also on convergence towards the true posterior.  We emphasize, however, that obtaining a high-fidelity approximation to the true posterior using ABC is not guaranteed, as this depends on a number of conditions to be met, including a careful selection of summary statistics. 
\cite{Silk2013} propose an adaptive approach for selecting the tolerance sequence at each iteration by estimating the threshold-acceptance rate curve (TAR curve), which is used to balance the amount of shrinkage of the tolerance with the acceptance rate.  This approach requires the estimation of the TAR curve at each iteration of the algorithm.  The naive, but computationally impractical approach to estimating the TAR curve (noted as such in \citealt{Silk2013}), is to simulate a Monte Carlo estimate of the acceptance rate at a range of different tolerances using the ABC forward model, which would have to be repeated at each iteration of the ABC algorithm.  Instead, they suggest a more practical method for estimating the TAR curve by building an approximation to the forward model (in their example, using a mixture of Gaussians and the unscented transform of \citealt{Julier:2000fk}). The TAR curve approach is able to avoid local optima values, but requires the extra step of building a fast approximation of the ABC data-generating model. Our proposed algorithm is similarly able to avoid local modes, but uses quantities that are directly available in the algorithm.  More details are presented in Section~\ref{sec:examples}.

After determining the sequence of tolerances, it is also necessary to determine when to stop a sequential ABC sampling algorithm.  An ABC algorithm is often stopped when either a desired (low) tolerance is achieved \citep{SissonEtAl2007} or after a fixed number of iterations $T$ \citep{BeaumontEtAl2009}. \cite{IshidaEtAl2015} showed that once the ABC posterior stabilizes, further reduction of the tolerance leads to low acceptance rates without meaningful improvement in the ABC approximation to the posterior. They stop the algorithm once the acceptance rate drops below a threshold set by the user. 

The first main contribution of this paper is to extend the ABC-PMC algorithm so that the quantile used to update the tolerance in each iteration, $q_t$, is automatically and efficiently selected, rather than being fixed in advance to a quantile that is used for each iteration. 
It is worth noticing that efficiency is not only a matter of having a high acceptance rate, as this can  be easily accomplished by using larger quantiles, but rather a balance between the acceptance rate and a suitable amount of shrinkage of the tolerance. 
Moreover the series of tolerances needs to be selected in such a way that the algorithm avoids getting stuck in local modes. As the second contribution, we develop an automatic stopping rule directly based on the behavior of the sequential ABC posterior.

The rest of the paper is organized as follows. In Section~\ref{sec:auto} the adaptive selection of $q_t$ for determining the tolerance sequence is presented along with the proposed stopping rule. Section~\ref{sec:examples} is dedicated to a simulation study to compare quantile-based selection of tolerances using ABC-PMC with the proposed procedure.  The final example considered uses real data on colonizations of the bacterium \textit{Streptococcus pneumoniae} \citep{numminen2013estimating}.  Concluding remarks are given in Section~\ref{sec:conclude}.

\section{Methodology} \label{sec:auto}

Using the same quantile to update the tolerance at each iteration can be computationally inefficient and results in the particle system getting stuck in local modes (see the example in Section \ref{sec:local_modes}).  
In this section we introduce a method for adaptively selecting the quantile such that each iteration has its own quantile, $q_t$, set based on the online performance of the algorithm. 

\subsection{Initial Sampling and Automatic Tolerance Selection Rule} 

In order to initialize the tolerance sequence we use the following approach.  
Let $N$ be the desired number of particles to approximate the posterior.
The initial tolerance $\epsilon_{1}$ can be adaptively selected by sampling $N\init=kN$ draws from the prior, for some $k \in \mathbb Z^+$ \citep{Cisewski-Kehe:2019aa}.  Then the $N$ particles of the $N\init$ total particles with the smallest distances are retained, and $\epsilon_1 = \max \left(d_1^{(1*)}, \ldots, d_1^{(N*)}\right)$, where $d_1^{(1*)}, \ldots, d_1^{(N*)}$ are the $N$ smallest distances of the $N\init$ particles sampled. 
This initialization procedure effectively selects a distance quantile for the first step by the selection of an appropriate $k$, but making this first step adaptive is easier than trying to guess a good $\epsilon_1$. Trying to specify a reasonable $\epsilon_1$ can be especially challenging when testing different summary statistics or distance functions because the scale of the distances can be different.
It is important to note that $k$ must be large enough to result in a satisfactory initial exploration of the parameter space, otherwise the algorithm might get stuck in local regions of the parameter space. This challenge also holds true in general for other ABC algorithms, including when $\epsilon_1$ is predefined (i.e. not set adaptively). Providing a general and suitable value for $k$ regardless of the problem that is considered is challenging, since this choice depends on a number of factors such as the definition of the prior distribution(s), the forward model and where relevant regions of the parameter space are (the latter being unknown). Therefore the parameter $k$ has to be suitably tuned by the user once the forward model and the prior distribution(s) have been defined. The problem of selecting $k$ is further discussed in Section~\ref{sec:examples}.

For the subsequent tolerances, $\epsilon_{2:T}$, 
 the general idea is to gauge the amount of shrinkage for iteration $t+1$ by determining the value of $\epsilon_{t+1}$ based on the amount of improvement between $\hat{\pi}_{\epsilon_{t-1}}$ and $\hat{\pi}_{\epsilon_{t}}$.
In particular, we can use the estimated ABC posteriors to select a quantile to update the tolerance for the next iteration, and adjust the next tolerance based on how slowly or rapidly the sequential ABC posteriors are changing. 
More specifically, after each iteration $t>1$, the following ratio can be estimated using the weighted particles:
\begin{equation}
\hat{c}_t =\sup_{\theta}\frac{{\hat{\pi}_{\epsilon_t}}(\theta)}{{\hat{\pi}_{\epsilon_{t-1}}(\theta)}}.
\label{eq:ter_eff}
\end{equation}
%
%
%
%
Since $\hat{\pi}_{\epsilon_{t-1}}(\theta)$ and $\hat{\pi}_{\epsilon_{t}}(\theta)$ from Eq. \eqref{eq:ter_eff} are both proper densities, they will be either exactly the same, making $\hat{c}_t=1$, or there must be a place where $\hat{\pi}_{\epsilon_{t}}(\theta) > \hat{\pi}_{\epsilon_{t-1}}(\theta)$, making $\hat{c}_t > 1$. 
Then the proposed quantile for iteration $t$ (in order to determine $\epsilon_{t+1}$) is
\begin{equation}
q_{t} =  \frac{1}{\hat{c}_{t}},
\label{eq:update_quantile}
\end{equation}
which varies between 0 and 1.  Small values of $q_t$ imply $q_{t-1}$ lead to a large improvement between $\hat{\pi}_{\epsilon_{t-1}}$ and $\hat{\pi}_{\epsilon_{t}}$, which then results in a larger percentage reduction of the tolerance for the coming iteration, $t+1$. On the other hand, once the ABC posterior stabilizes, $q_t$ tends to $1$ as $\hat{\pi}_{\epsilon_{t-1}}$ and $\hat{\pi}_{\epsilon_{t}}$ become more similar.
The form of Eq. \eqref{eq:update_quantile} was motivated by the Accept-Reject (A/R) algorithm \citep{andrieu2003introduction, robert2013monte}. The A/R algorithm has a target distribution, a proposal distribution, and a rule to decide whether or not an element coming from the proposal distribution should be accepted as an element coming from the target distribution. 
If the form of the ABC posterior distribution was known, A/R sampling would work as follows.   A candidate, $\theta^*$, would be proposed from $\hat{\pi}_{\epsilon_{t-1}}(\theta | y_{\obs})$, and would be accepted with probability $\frac{\hat{\pi}_{\epsilon_t}(\theta^* | y_{\obs})}{c \cdot \hat{\pi}_{\epsilon^*_{t-1}}(\theta | y_{\obs})}$, where $c \in (1, \infty)$ is a positive real constant number selected such that  $\hat{\pi}_{\epsilon_t}(\theta | y_{\obs}) \le c \cdot \hat{\pi}_{\epsilon_{t-1}}(\theta | y_{\obs})$ \citep{robert2013monte}.
In A/R sampling, the unconditional acceptance probability is $\frac{1}{c}$ \citep{hesterberg1988advances}.
The constant $c$ acts as a proxy for the difference between the proposal and the target distributions (e.g., if they are the same distribution, then $c = 1$ and all proposals would be accepted).

The ABC algorithm does not follow the A/R sampling scheme, but the notion of $1/c$ relating to the sampling efficiency in the A/R algorithm inspired the proposed adaptive tolerance selection idea. For some future iteration, say iteration $t+1$, the ABC posterior distribution is unknown so the previous two ABC posteriors at iterations $t-1$ and $t$ are used as the proposal and target distributions, respectively, so that $\hat{c}_t$ can be computed in Eq.~\eqref{eq:ter_eff}.  
If there was a substantial change between $\hat{\pi}_{\epsilon_{t-1}}$ and $\hat{\pi}_{\epsilon_t}$, then $\hat{c}_t$ would be larger resulting in a smaller quantile, $q_t$, for specifying $\epsilon_{t+1}$.
As $\hat{\pi}_{\epsilon_{t-1}}$ and $\hat{\pi}_{\epsilon_t}$ become more similar, larger quantiles $q_t$ are assigned.
The proposed form of $c_t$ allows the tolerance selection to be based on changes in the ABC posterior from the previous iteration where substantial changes between iterations $t-1$ and $t$ result in a substantial decrease in the proposed tolerance for $\epsilon_{t+1}$.  This continues until a substantial decrease in the proposed tolerance does not result in a substantial change in the ABC posterior, at which point the amount of shrinkage in the tolerance becomes smaller.


We found the rule based on Eq. \eqref{eq:update_quantile} to work well empirically.  One challenge with a theoretical evaluation of the proposed algorithm, and other algorithms designed to optimize the tolerance shrinkage and acceptance rate, is that the acceptance rate depends on the forward simulation model.  In general ABC settings, the forward simulation model does not have a closed-form expression.



\begin{figure}[htbp]
\centering
\includegraphics[width=.8\textwidth]{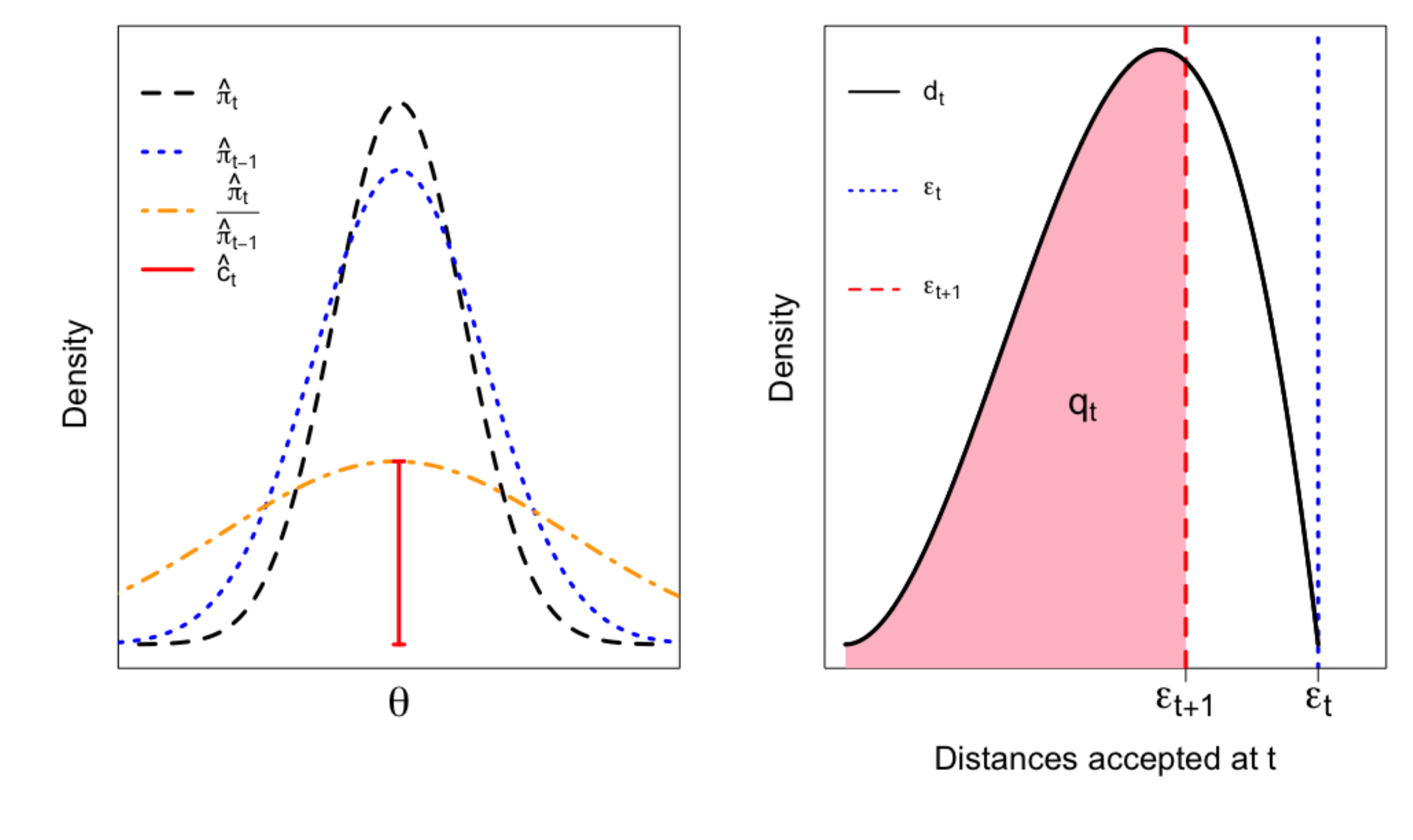}
\caption{Illustration of the selection of $q_t$. (left) The proposal distribution  ABC posterior $\hat{\pi}_{t-1}$, the resulting ABC posterior $\hat{\pi}_t$ and their ratio $\frac{\hat{\pi}_t}{\hat{\pi}_{t-1}}$, with  $\hat{c}_t$ defined according to Eq. \eqref{eq:ter_eff} and used for setting $q_t$, as defined in Eq. \eqref{eq:update_quantile}. (right)  The (arbitrary) distribution of distances is from the accepted distances at iteration $t$, $\{d_t^{(J)}\}_{J = 1}^N$, with $\epsilon_t$ being the largest possible value.  The next iteration's tolerance, $\epsilon_{t+1}$, is set as the $q_t$ quantile of $\{d_t^{(J)}\}_{J = 1}^N$.
}
\label{fig:justification}
\end{figure}

An illustration of the proposed quantile selection procedure is provided in Figure~\ref{fig:justification}.
If $\hat{\pi}_{t-1}$ was used as the proposal for iteration $t+1$ (instead of $\hat{\pi}_t$), then $q_t$ would be the percentage decrease in the acceptance rate from iteration $t$, \textit{i.e.} if $\text{acc}_t$ is the acceptance rate for iteration $t$, then $\text{acc}_{t+1}$ would be approximately $q_t \times \text{acc}_t$.  However, we are not proposing from $\hat{\pi}_{t-1}$, but rather $\hat{\pi}_t$ so the \emph{decrease} in the acceptance rate is mitigated by the \emph{improvement} in the proposed particles from iteration $t$.  When there is a large improvement in the ABC posterior from $\hat{\pi}_{t-1}$ to $\hat{\pi}_t$, then $q_t$ is smaller, allowing for a larger drop in the tolerance. This larger percentage drop in tolerance does not result in an equal percentage drop in acceptance rate because the new proposal distribution, $\hat{\pi}_t$, is better than $\hat{\pi}_{t-1}$. Conversely, if $\hat{\pi}_{t-1}$ is close to $\hat{\pi}_t$, then the improvement in the ABC posterior is not enough to allow for a large decrease in the acceptance rate and consequently $q_t$ is closer to 1.

The evaluation of Eq. \eqref{eq:ter_eff} relies on the calculation of the ratio between the (possibly multidimensional) density functions, defined here as $r$. A naive solution would be to separately calculate the density for $\hat{\pi}_t$ and $\hat{\pi}_{t-1}$ using some Kernel Density Estimate (KDE) method (see \citealt{silverman2018density} for a review), and then estimate the ratio from those estimates. Then, the supremum of the previously calculated ratio can be obtained, for example, through an optimization procedure that computes the density over a grid of values. However, this is not a reliable solution, in particular for high-dimensional cases for which division by an estimated quantity can magnify the estimation error \citep{sugiyama2008direct}. In order to address the problem of properly estimating $r$ with $\hat{r}$, and therefore solving Eq. \eqref{eq:ter_eff}, alternatives to the KDE solution are available, such as ratio estimation methods (REM) \citep{sugiyama2012density}. The main advantage of using REM is that the calculation of the desired ratio does not include density estimation, which would involve dividing by an estimated KDE. Additionally, when using a KDE, kernel and bandwidth need to be selected, which can affect the result. Poorly estimating the density of the denominator of $r$, in particular, can potentially increase the error of the estimated ratio \citep{sugiyama2010density}.
There are several different REM frameworks (e.g. \citealt{bickel2007discriminative,gretton2009covariate,sugiyama2008direct,sugiyama2010density}), but we use the ratio matching approach of \cite{sugiyama2008direct} discussed in more detail next.

In order to introduce the REM framework, consider $\theta \in \mathbb R^p$ and two generic samples $\{\theta_i^{L}\}_{i=1}^{L}$ and $\{\theta_j^{M}\}_{j=1}^{M}$, where \textit{L} and \textit{M} are the sample sizes for the first and the second sample, respectively. The sample $\{\theta_i^{L}\}_{i=1}^{L}$ has as corresponding density $p_{L}(\theta)$, while the sample $\{\theta_i^{M}\}_{i=1}^{M}$ has as corresponding density $p_{M}(\theta)$. The density ratio $r(\theta)$ can be defined as $r(\theta)=\frac{p_{L}(\theta)}{p_{M}(\theta)}$.
The basic idea of the ratio matching approach is to match a density ratio model $\hat{r}(\theta)$ with the true density ratio $r(\theta)$ under some divergence \citep{sugiyama2010density}. Several divergences can be used to compare $\hat{r}(\theta)$ with $r(\theta)$.  A common divergence is the Bregman divergence \citep{bregman1967relaxation}, along with some of its related divergences such has the unnormalized Kullback-Leibler divergence and the squared distance. In particular, the unnormalized Kullback-Leibler divergence minimizes the divergence between $p_{L}(\theta)$ and $\hat{p}_{L}(\theta)=\hat{r}(\theta) p_{M}(\theta)$ by means of the following criterion:
\begin{equation}
\min_{\hat{r}} \int p_{L}(\theta) \log{\frac{p_{L}(\theta)}{\hat{r}(\theta)p_{M}(\theta)}} d\theta.
 \label{eq:KL}
\end{equation}
By decomposing the Kullback-Leibler divergence defined in  Eq. \eqref{eq:KL}, $\hat{r}(\theta)$ can be estimated by solving the objective function $\max_{\hat{r}} \int p_{L}(\theta) \log{\hat{r}(\theta)} d\theta$ \citep{hido2011statistical, sugiyama2010density}. Further details on the unnormalized Kullback-Leibler divergence and on other REM approaches are found in \cite{sugiyama2012density}. As pointed out by \cite{sugiyama2010density}, a further non-negligible advantage of using REM, and in particular the ratio matching approach, is the applicability of gradient-based algorithms and quasi--Newton methods for optimization over $\hat{r}(x)$. 

In the analyses of the present work we use the ratio matching approach and the Kullback--Leibler importance estimation procedure (KLIEP) \citep{hido2011statistical, sugiyama2010density,sugiyama2008direct} in order to estimate, at the end of each iteration $t$, the ratio of densities defined in Eq. \eqref{eq:ter_eff}.  Recall that the densities involved in Eq. \eqref{eq:ter_eff} are  $\hat{\pi}_{\epsilon_t}(\theta)$ and $\hat{\pi}_{\epsilon_{t-1}}(\theta)$. Once the ratio between $\hat{\pi}_{\epsilon_t}(\theta)$ and $\hat{\pi}_{\epsilon_{t-1}}(\theta)$ has been estimated, the supremum of Eq. \eqref{eq:ter_eff} is calculated by using an optimizer over the parameter space, such as the one proposed by \cite{brent2013algorithms}. The quantile used to reduce the tolerance for the coming iteration is finally retrieved by using Eq. \eqref{eq:update_quantile}. The steps discussed  above are performed at the end of each iteration as long as the stopping rule, defined in Eq. \eqref{eq:stoppingruleEfficiency} and discussed below, is not satisfied.
Estimation of $\hat{r}$ is carried out by using the densratio package\footnote{https://github.com/hoxo-m/densratio}, which is freely available in the R software \citep{Rcite}.

The acceptance rate is also useful for evaluating the computational burden of the ABC-PMC algorithm, defined as:
\begin{equation}
\text{acc}_t = \frac{N}{\D_t},
\label{eq:real_eff}
\end{equation}
where $\D_t$ is the number of draws done at iteration $t$ in order to produce $N$ accepted values. Eq. \eqref{eq:real_eff} generally decreases with each iteration because as the tolerance decreases, the number of elements $\D_{t}$ required to get $N$ accepted particles generally increases \citep{lintusaari2017fundamentals}.

\subsection{Stopping Rule} 

There are several published ideas in the literature on how to determine the number of iterations in an ABC-PMC algorithm. Often one picks some $T$ based on the computational resources available, but this can be needlessly inefficient. \cite{IshidaEtAl2015} proposed to stop the algorithm once the acceptance rate is smaller than some specified, fixed tolerance. The proposed stopping rule is directly based on the estimated sequential ABC posterior distributions, which avoids unnecessary additional iterations of the algorithm.

The ABC--PMC algorithm produces a sequence of $T$ posterior distributions, $\hat{\pi}_{\epsilon_t}$, where $\epsilon_t$ identifies the tolerance used in iteration $t$, with $t = 1, \dots, T$ and $\epsilon_1 > \epsilon_2 > \dots > \epsilon_T$. When defining a stopping rule, it turns out that Eq. \eqref{eq:update_quantile} can be used not only to adaptively selecting the quantile used to reduce the tolerance across the iterations, but also to indicate when to stop the procedure once the sequential ABC posterior stops changing significantly.

The series of quantiles defined through Eq. \eqref{eq:update_quantile} generally increases as the tolerance decreases. In particular, since the quantile used to reduce the tolerance is based on the online performance of the ABC posterior distribution, once the ABC posterior has stabilized, $q_{t} \approx 1$. This follows directly from Eq. \eqref{eq:ter_eff} because once the ABC posterior has stabilized $\hat{c}_t \approx 1$, and further reductions of the tolerance (i.e. additional iterations) do not necessarily lead to an improvement by the ABC posterior distribution.
In other words, once the ABC posterior stabilizes, the series of the quantiles defined through Eq. \eqref{eq:update_quantile} stops increasing and the upper bound of $1$ implies that no further reduction will improve the ABC posterior distribution. This leads to an automatic and simple stopping rule, which is employed starting from the third iteration, \textit{i.e.} once the transformation kernel has been used twice to avoid premature stopping. Our algorithm is stopped at time $t$ when
\begin{equation}
q_{t} > 0.99 \text{ for } t\geq3.
\label{eq:stoppingruleEfficiency}
\end{equation}
Hence, the algorithm is stopped once the quantile used to reduce the tolerance suggests that further reduction is not necessary since the ABC posterior has stabilized. 

Using Eq. \eqref{eq:update_quantile} as an automatic rule to shrink the tolerance and Eq. \eqref{eq:stoppingruleEfficiency} as the stopping rule, the ABC-PMC algorithm is stopped once additional iterations with smaller tolerances do not lead to significant changes in the ABC posterior. \footnote{The desired sample size $N$ has an impact on the evaluation of Eq. \eqref{eq:stoppingruleEfficiency}. This problem arises also in the classical MCMC analysis when determining the length of the MCMC chain \citep{GelmanEtAl2014}.  An $N$ that is too small leads to more variability of the estimated posterior in Eq. \eqref{eq:stoppingruleEfficiency}, which could lead to the algorithm stopping prematurely.}

\section{Illustrative Examples} \label{sec:examples}

Next we provide a comparison between the original ABC-PMC algorithm and our extension proposed in Section~\ref{sec:auto}, the aABC-PMC, by using three examples. 
In the first example the Gaussian mixture model by \cite{SissonEtAl2007} is used in order to demonstrate the computational efficiency of the proposed aABC-PMC procedure. 
Then the aABC-PMC algorithm is used for a model from \cite{Silk2013}, which has local modes, in order to illustrate how the proposed automatic tolerance selector is able to avoid getting stuck in local regions of the parameter space.
%
%
The final example, originally presented in \cite{numminen2013estimating}, uses data on colonizations of the bacterium \textit{Streptococcus pneumoniae} and represents a computationally expensive forward model.
Expensive forward models are a challenge for ABC methods because the computational cost can be prohibitive for practical applications, and in these cases selecting an appropriate sequence of tolerances is crucial.
A fourth example, the Lotka--Volterra model by \cite{TonyEtAl}, is presented in the Appendix A of the Supplementary Material.

In order to compare the proposed procedure with the original ABC-PMC algorithm, both the computational time and the total number of draws until the stopping criterion is satisfied are considered. The Hellinger distance is used for evaluating the similarity between the 1--dimensional marginal ABC posterior distributions at the final iteration, $\hat{\pi}_{\epsilon_T}$, and a benchmark, ${\pi}_{\text{true}}$, which is defined as:
\begin{equation}
H(\hat{\pi}_{\epsilon_T},{\pi}_{\text{true}}) = \left(\int \left(\sqrt{\hat{\pi}_{\epsilon_T}(y)}-\sqrt{{\pi}_{\text{true}}(y)} \right)^2 dy \right)^{\frac{1}{2}}.
\label{eq:hellinger}
\end{equation}
The benchmark, ${\pi}_{\text{true}}$, is the true posterior distribution if it is available in closed form, which is the case in the first two presented examples (see Sections~\ref{sec:gmm} and \ref{sec:local_modes}). In the final  example, since the true posterior distribution is not available, the ABC posteriors from \cite{numminen2013estimating}, are used as benchmarks (see Section~\ref{sec:daycarel}).

In order to estimate the 1--dimensional marginal ABC posterior distributions from the samples and their corresponding importance weights, a KDE \citep{silverman2018density} is used with a Gaussian kernel and a smoothing bandwidth parameter $h$. The bandwidth is selected using Silverman's rule--of--thumb \citep{silverman1986density}.

Finally, unless otherwise noted, the number of particles in the ABC procedures is set to $N=1,000$.

\subsection{Gaussian Mixture Model} \label{sec:gmm}

The first application of the aABC-PMC is an example from \cite{SissonEtAl2007}, which is also analyzed by \citet{BeaumontEtAl2009}.  
It is a Gaussian mixture model with two Gaussian components with known variances and mixture weights, but an unknown common mean, $f(y \mid \theta)=0.5 \mathcal{N}(\theta,1) + 0.5 \mathcal{N}(\theta,0.01)$ and prior $\pi(\theta) \sim \text{Unif}(-10,10)$. With a single observation $y\obs=0$, the true posterior distribution is

\begin{equation}
\pi(\theta \mid y\obs) \sim 0.5 \mathcal{N}(0,1) + 0.5 \mathcal{N}(0,0.01).
\label{eq:posteriorMixture}
\end{equation}

\begin{figure}[htbp]
\centering
\subfloat[][{Final ABC posteriors}.]
{\includegraphics[height=.43\columnwidth]{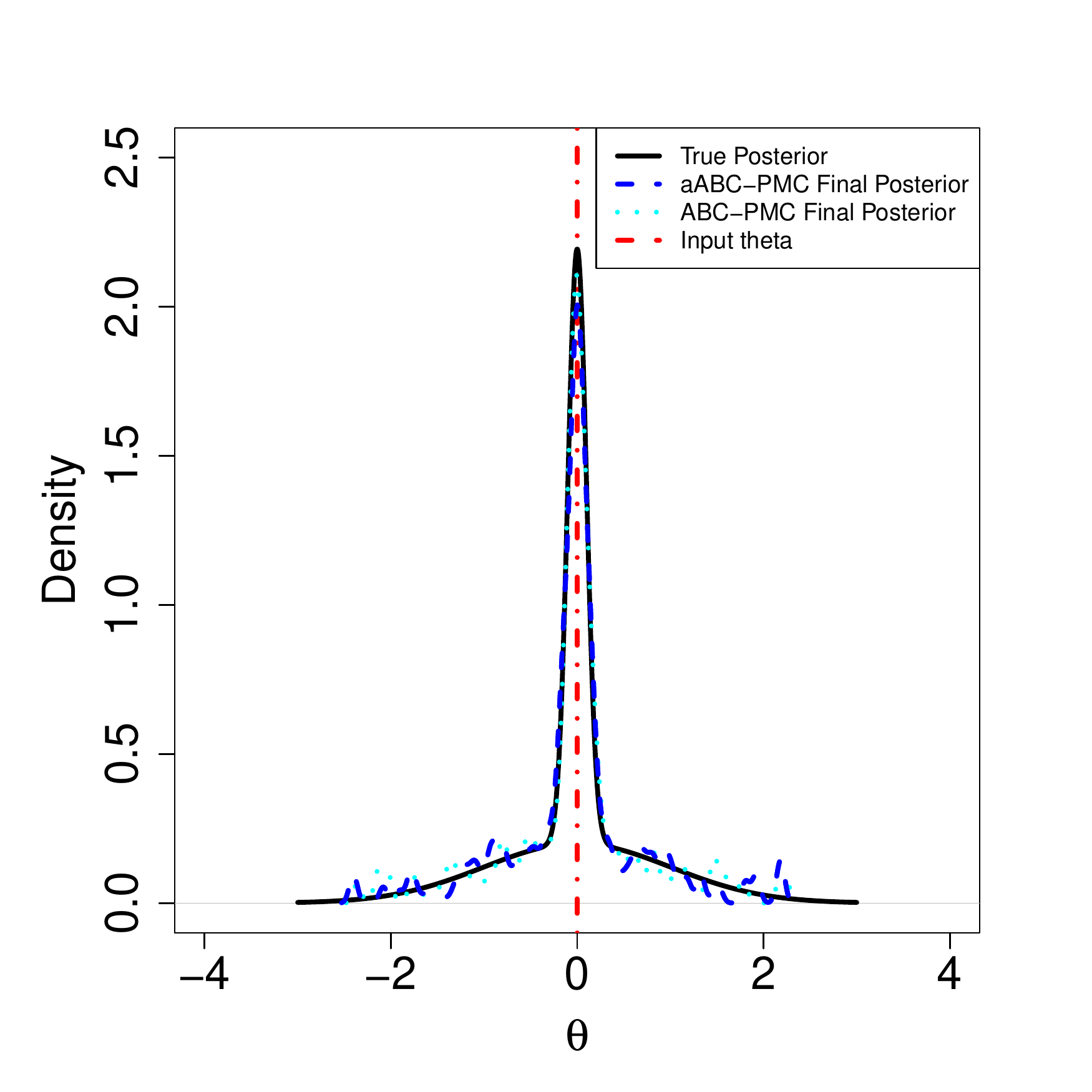}
\label{fig:abcpostSisson}} \quad
\subfloat[][{aABC-PMC quantities}.]
{\includegraphics[height=.43\columnwidth]{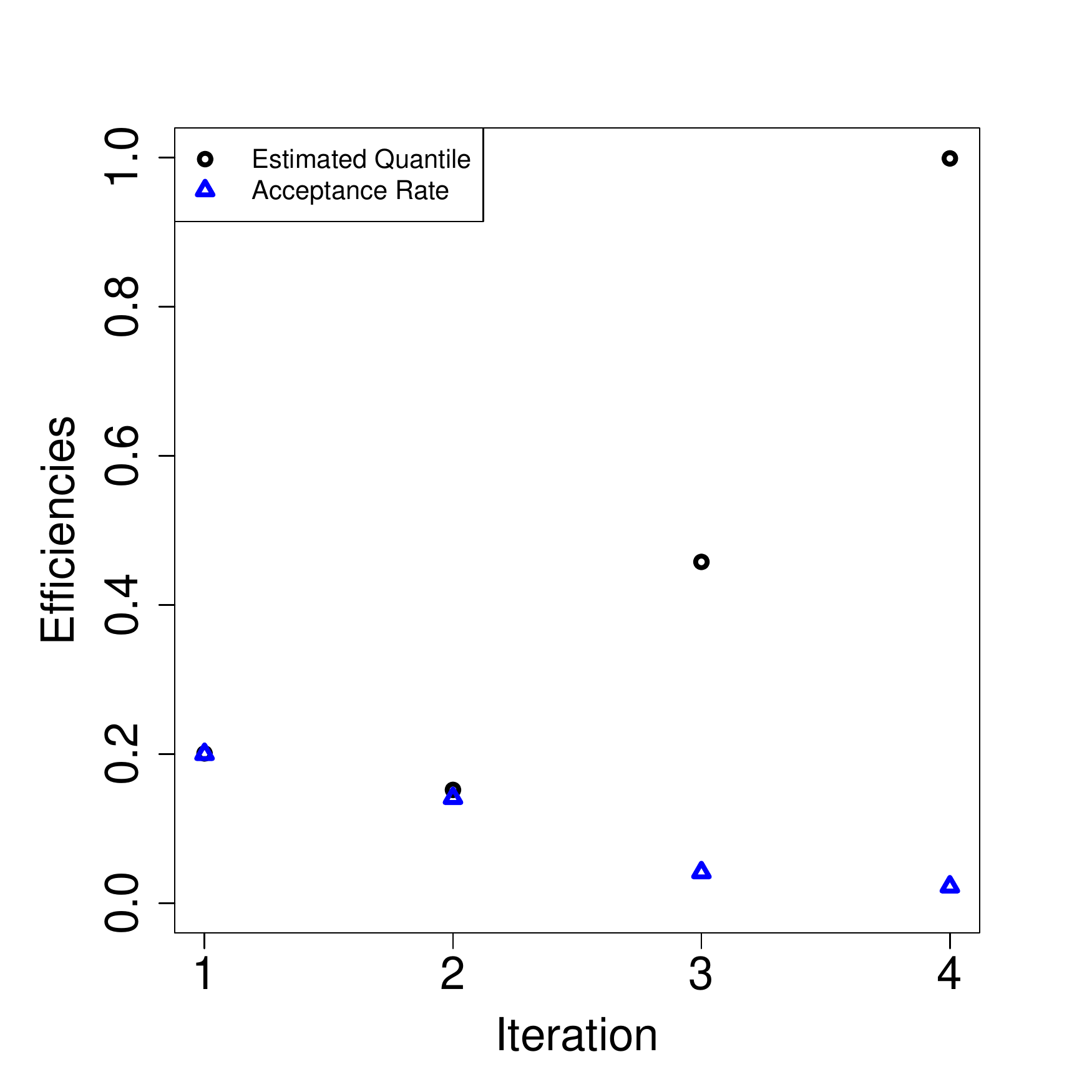}
\label{fig:quantilesSisson}} \quad
\caption{Gaussian mixture model example. (a) ABC-PMC and aABC-PMC final posterior distributions and (b) sequential quantities computed for the aABC-PMC method.  
The $q_t$'s (black circles) generally increase through the iterations until the ABC posterior has stabilized. The acceptance rate (blue triangles) decreases throughout the iterations, which is why it is desirable to stop the algorithm once the ABC posterior has stabilized.}
\label{fig:SissonEx}
\end{figure}

For consistency with the results of  \cite{SissonEtAl2007} and \citet{BeaumontEtAl2009}, the distance function used is $\rho \left(y\obs, y\pro\right)= |y\obs - y\pro|$, $N=1,000$, and a Gaussian kernel for resampling the particles is used. Both \cite{SissonEtAl2007} and \citet{BeaumontEtAl2009} manually define the series of tolerances. In particular, \cite{SissonEtAl2007} carry out $T=10$ iterations with a fixed series of tolerances $\epsilon_{1:10}$ displayed in Table \ref{tab:Sisson10steps}. 
%
To evaluate the reliability of the aABC-PMC, a comparison with the ABC-PMC is done both in terms of computational time and total number of draws. The results of the analysis are shown in Table \ref{tab:Sisson10steps} and are based on $21$ independent runs with the same dataset, $y\obs=0$. 
The table includes the values for the run that produced the median number of total draws.
The aABC-PMC outperforms ABC-PMC in the terms of total draws (81,230 vs 1,421,283) and a faster computational time (88 seconds vs 243 seconds). The final ABC posteriors for each method are displayed in Figure~\ref{fig:abcpostSisson}.  Though the aABC-PMC method is computationally more efficient than the ABC-PMC approach, the final ABC posteriors are very similar. This suggests that after a suitable tolerance is achieved, decreasing the tolerance further does not necessarily lead to a better approximation of the posterior distribution. 
%

From Table \ref{tab:Sisson10steps}, we note that the final tolerance for \cite{SissonEtAl2007} is $\epsilon_{10}=0.0025$ ($H\dist=0.20$) while the automatic stopping rule of aABC-PMC leads to $4$ iterations with a final tolerance of $\epsilon_{4} = 0.035$ ($H\dist=0.20$). 
In Figure~\ref{fig:quantilesSisson}, the $q_t$'s retrieved by using Eq.~\eqref{eq:update_quantile} are displayed (black circles), which increase until the final iteration, while the acceptance rate (blue triangles) decreases.
Neglecting to stop the algorithm once the ABC posterior has stabilized can be inefficient since the number of draws needed in order to complete further iterations can drastically increase, as evidenced by the increasing $\D_t$ for later iterations displayed in Table~\ref{tab:Sisson10steps}.

\begin{table*}\centering
\ra{1.2}
\begin{tabular}{@{}rrrrcrrrcrrr@{}}\toprule
& \multicolumn{4}{c}{Sisson et al. (2007)} & \multicolumn{4}{c}{aABC-PMC}\\
\cmidrule{1-5} \cmidrule{6-10}
$t$ & $\epsilon_{t}$ & $\D_{t}$& $H\dist$ &&$t$ & $\epsilon_{t}$ & $q_{t}$  & $\D_{t}$& $H\dist$ \\\midrule
1 & 1.000 & 2,595 & 0.34 && 1 & 1.96 &  & 5,000 & 0.39\\
2 & 0.5013 & 8,284 & 0.29 && 2 & 0.45 & 0.20 & 7,095 & 0.29\\
3 & 0.2519 & 8,341 & 0.26 && 3 & 0.072 & 0.15 & 24,216 & 0.22\\
4 & 0.1272 & 7,432 & 0.24 && 4 & 0.035 & 0.45 & 44,919 & 0.20\\
5 & 0.0648 & 10,031 & 0.23 &&  &  &  &  &  \\
6 & 0.0337 & 17,056 & 0.20 &&  &  &  &  & \\
7 & 0.0181 & 34,178 & 0.21 && & &  & & \\
8 & 0.0102 & 72,704 & 0.20 &&  & &  & & \\ 
9 & 0.0064 & 171,656 & 0.19  && &  &  & & \\
10 & 0.0025 & 1,089,006 & 0.20 && &  &  & & \\
\bottomrule
Total &  & 1,421,283 & &&  &  & & 81,230 \\
\end{tabular}
\caption{Gaussian mixture model. The number of draws needed in each iteration to reach $N=1,000$ accepted values for the ABC-PMC and the aABC-PMC algorithm. (The displayed results were obtained by running the procedure 21 times and using the run that produced the median number of total draws.)
For the aABC-PMC algorithm, the quantile  automatically selected through the iterations is displayed under $q_{t}$. The procedure stopped once the quantile $q_5 = 0.999$ was proposed. For the ABC--PMC algorithm a total of $1,421,283$ ($243$ sec.) draws were required, while our aABC-PMC takes $81,230$ ($88$ sec.) draws overall.}
\label{tab:Sisson10steps}
\end{table*}

Next, we show the behavior of the aABC-PMC algorithm for different choices of the number of proposed values from the prior distribution at the first iteration of the procedure.  
Initial particle sample sizes, $N\init$, of N, 2N, 5N, and 10N are considered (with $N = 1,000$), and the results are displayed in Table \ref{tab:GMM_Efficiency_Ninit}. The initial particle sample size that seems to best balance the total number of draws and the time required to satisfy the stopping rule in this example is $5N$, with similar final ABC posterior distributions based on $H\dist$ (see Table \ref{tab:GMM_Efficiency_Ninit}); the posteriors are displayed in Figure \ref{fig:GMM_Efficiency_Ninit_Post}.

\begin{table}
\begin{minipage}[c]{0.50\linewidth}
\centering
\resizebox{\columnwidth}{!}{%
\scalebox{1}{%
\begin{tabular}{@{}rrrrcrrrcrrr@{}}\toprule
\cmidrule{1-7}
 & T & $\text{D}_{t}$ & $\epsilon_{1}$ & $\epsilon_{T}$ & time (sec) & $H\dist$  \\
N & 14 & 276,885 & 11.54 & 0.035 & 208 & 0.23  \\
2N & 10 & 109,720 & 4.97 & 0.077 & 150  & 0.29  \\
5N & 4 & 81,230 & 1.96 & 0.035 & 88  & 0.20  \\
10N & 4 & 90,194 & 1.00 & 0.059 & 105  & 0.17  \\
\bottomrule
\end{tabular}}
}
\caption{aABC-PMC algorithm with different choices for $N\init$ $(N, 2N, 5N, 10N)$ for the Gaussian mixture model example.}
\label{tab:GMM_Efficiency_Ninit}
\end{minipage}\hfill
\begin{minipage}[c]{0.48\linewidth}
\centering
\includegraphics[width=.95\columnwidth]{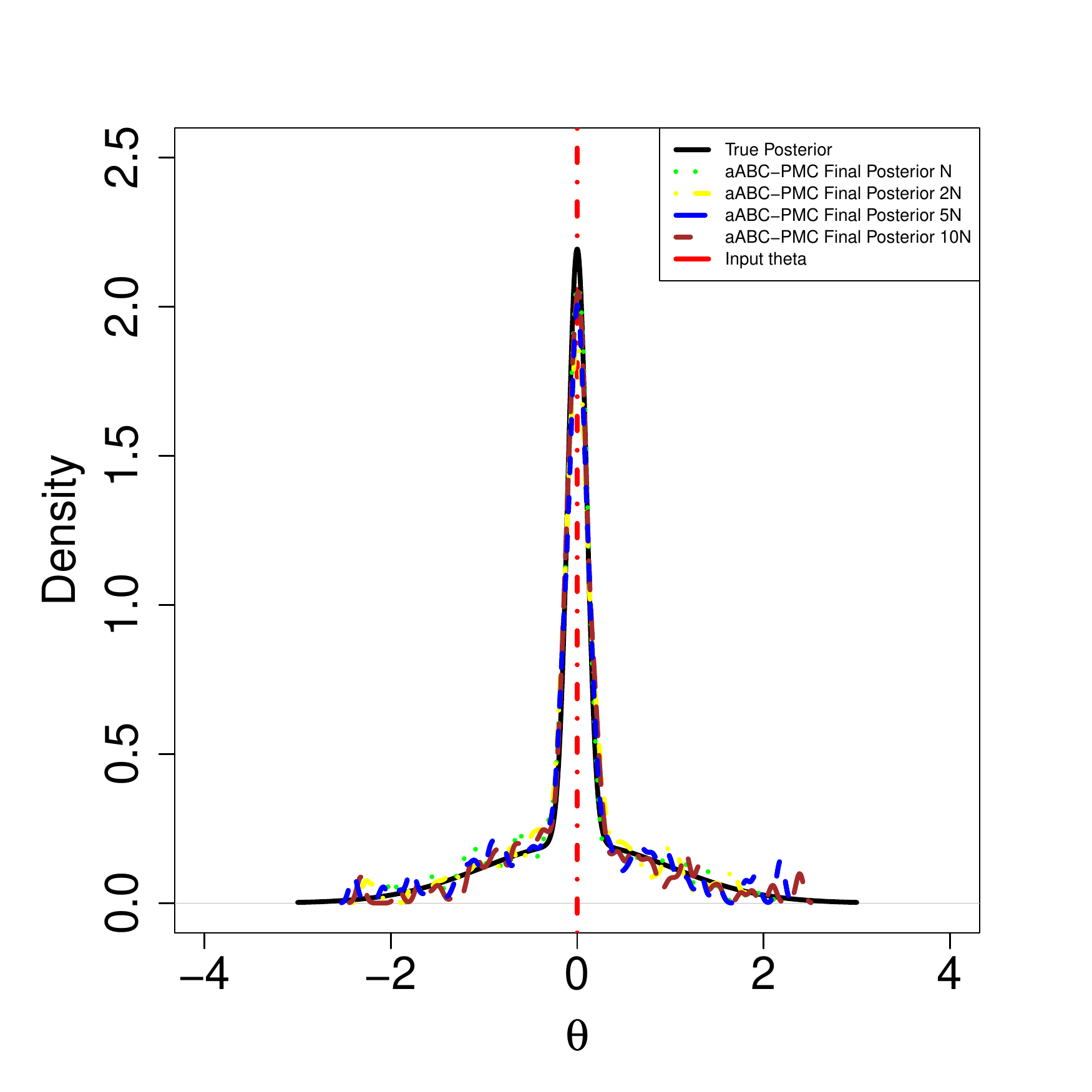}
\captionof{figure}{\hspace{-.1cm}: aABC-PMC posteriors with different choices for $N\init$ $(N, 2N, 5N, 10N)$ for the Gaussian mixture model example.
}
\label{fig:GMM_Efficiency_Ninit_Post}
\end{minipage}
\end{table}

Using the EasyABC {\tt R} package\footnote{\url{https://cran.r-project.org/web/packages/EasyABC}.} we carried out the same analysis for the ABC-SMC algorithm by \cite{DelMoralEtAl2012}. The ABC--SMC algorithm by \cite{DelMoralEtAl2012} is discussed in Section \ref{sec:daycarel}. For each initial particle sample size, $N\init$, of N, 2N, 5N, and 10N, 21 independent runs with the same dataset are performed and the runs that produced the median number of total draws are compared to the corresponding run obtained by our adaptive approach. 
Our choices for setting the parameters required by the ABC-SMC algorithm (see Sec. 3.4) are: $N=1,000$, $\epsilon=0.035$, $\alpha=0.5$, $M=1$ and $\text{nb}_{\text{threshold}}=N/2$. We note that for the last three parameters, the default values are used, according to the suggestions by \cite{DelMoralEtAl2012}. The results of the analysis are summarized in Table \ref{tab:ABCSMC_GMM_Efficiency_Ninit} and the corresponding posterior distributions are displayed in Fig. \ref{fig:SissonEx_ABCSMC}. For all four $N\init$ values considered, the final tolerances returned by the ABC-SMC algorithm are comparable with the one obtained by our approach with $N\init = 5N$ ($\epsilon_4=0.035$). 
However the corresponding ABC-SMC posterior distributions do not match the true posterior distribution as well as our proposed approach.  In particular, the ABC--SMC algorithm does not seem to capture the (low) variance coming from the second component of the Gaussian mixture model. 
Similar results were also obtained by the ABC--SMC sampler proposed in \cite{Bonassi2015}. 
A further comparison when using the ABC--SMC algorithm by \cite{DelMoralEtAl2012} is available in the Appendix B of the Supplementary Material.

\begin{table}
\begin{minipage}[c]{0.50\linewidth}
\centering
\resizebox{\columnwidth}{!}{%
\scalebox{1}{%
\begin{tabular}{@{}rrrrcrrrcrrr@{}}\toprule
\cmidrule{1-5}
 &  $\text{D}_{t}$ & time (sec)  & $H\dist$ &  $\epsilon_{t}$\ \\
N  & 25,023 & 17  & 0.35 & 0.037 \\
2N  & 47,192  & 54    & 0.37 & 0.031 \\
5N & 124,890 & 322   & 0.33 & 0.031 \\
10N  & 249,696   &  1254 & 0.34 & 0.03  \\
\bottomrule
\end{tabular}}
}
\caption{ABC-SMC algorithm with different choices for $N\init$ $(N, 2N, 5N, 10N)$ for the Gaussian mixture model example. 
}
\label{tab:ABCSMC_GMM_Efficiency_Ninit}
\end{minipage}\hfill
\begin{minipage}[c]{0.48\linewidth}
\centering
\includegraphics[width=.95\columnwidth]{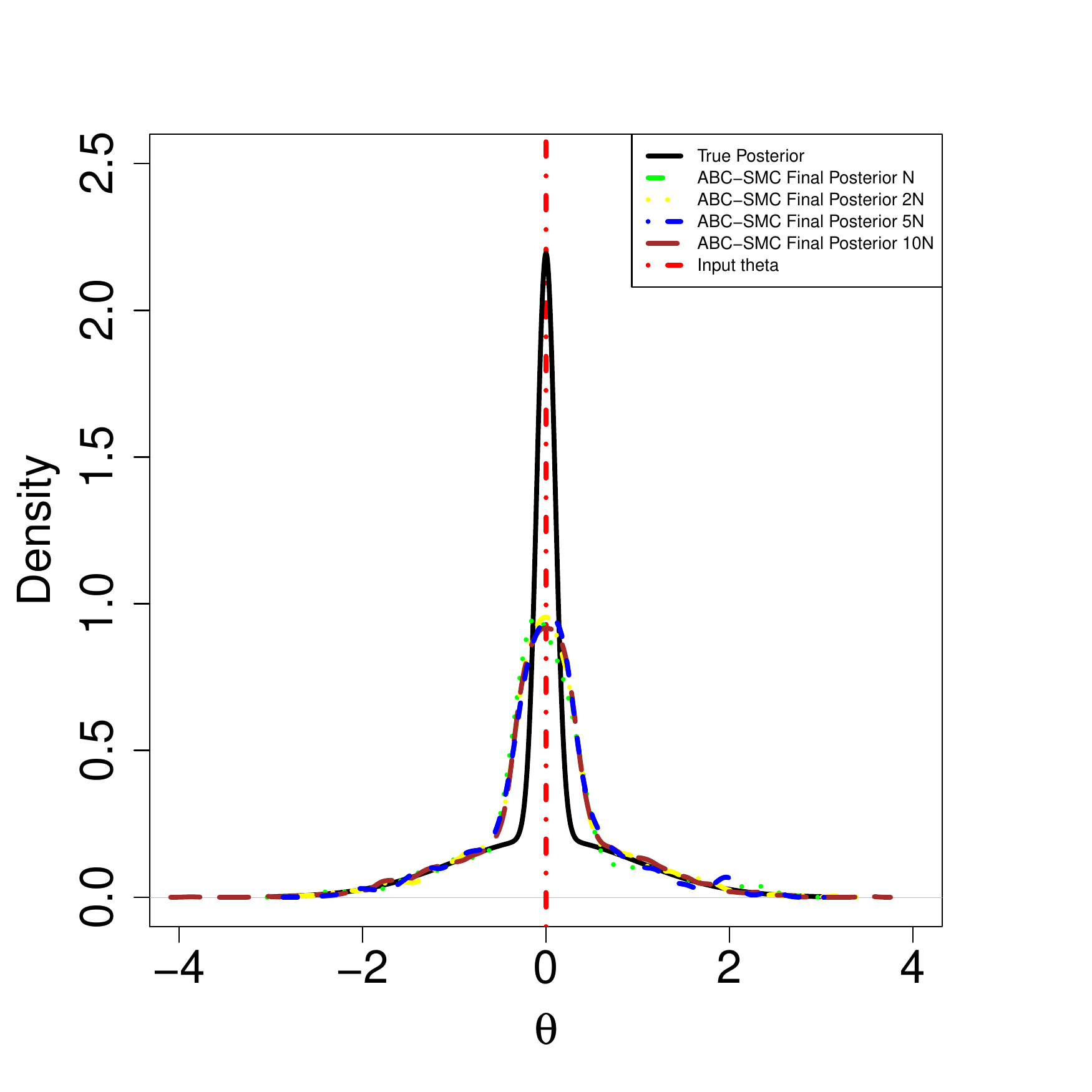}
\captionof{figure}{\hspace{-.1cm}: ABC-SMC final posterior distributions with different choices for $N\init$ $(N, 2N, 5N, 10N)$ for the Gaussian mixture model example. 
}
\label{fig:SissonEx_ABCSMC}
\end{minipage}
\end{table}

\subsection{Presence of a Local Mode} \label{sec:local_modes}
The sequence of tolerances has an impact not only on the computational efficiency of the algorithm, but also on its ability to find the true posterior \citep{Silk2013}, noting again that convergence to the true posterior using ABC is not guaranteed. 
To demonstrate the performance of aABC-PMC in the presence of local modes, we consider an example proposed in \cite{Silk2013}.
The (deterministic) forward model is $g(\theta)=(\theta-10)^2-100\exp(-100(\theta-3)^2)$.  The input value is set to $\theta=3$ leading to a single observation $y\obs=-51$. The true posterior distribution is a Dirac function at $3$. The specifications for the distance function ($L^1$ norm), the prior distribution (a normal distribution with mean of $10$ and variance of $10$), and the desired number of particles ($N=1,000$) are taken from \cite{Silk2013}.

\begin{figure}[htbp]
\centering
{\includegraphics[width = .65\textwidth]{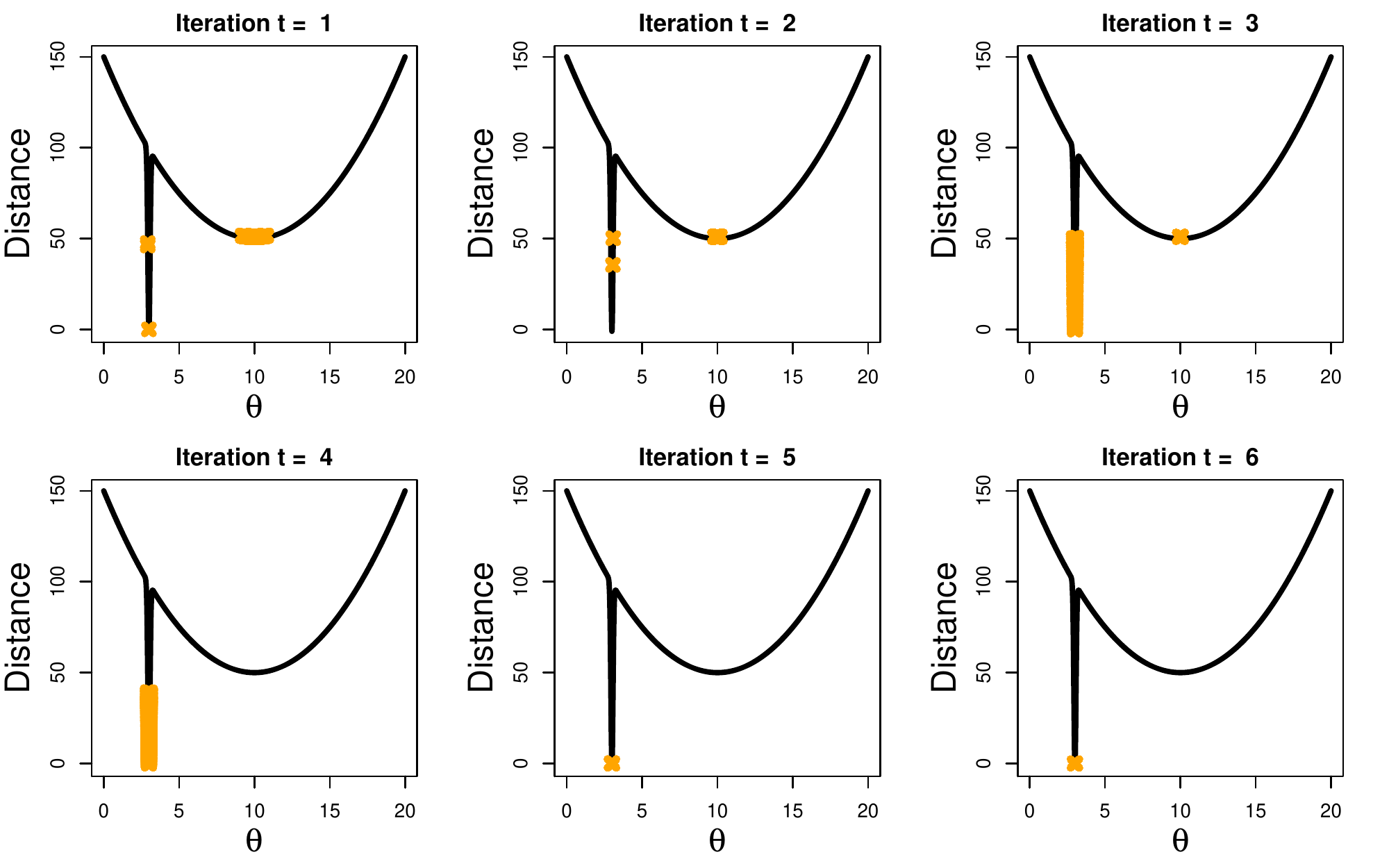}} \quad
\caption{Example from \cite{Silk2013} to investigate the performance of the proposed aABC-PMC in the presence of a local optimal value.  The accepted $\theta$ are plotted as orange x's against the corresponding distance by iteration.
}
\label{fig:LocalMinEx}
\end{figure}

Figure~\ref{fig:LocalMinEx} displays the locations of the accepted particles (orange x's) against the distances for a range of $\theta$'s, which highlights the challenge for ABC with this model. There is a local minimum distance around $\theta=10$, but the global minimum distance occurs at the true value of $\theta = 3$.  Initial steps of the ABC algorithm will find the local minimum, but the algorithm can easily get stuck around $\theta = 10$ if the sequential tolerances are not selected carefully.
The series of plots in Figure \ref{fig:LocalMinEx} shows the behavior of the aABC-PMC algorithm by focusing on the values for $\theta$ that were accepted (orange x's).
After $6$ iterations, the aABC-PMC algorithm has found the global minimum distance around the true $\theta$. The results of the analysis, based on $21$ independent runs, are summarized in Table \ref{tab:Silk}, where $384,347$ total particles were used by the proposed aABC-PMC algorithm. The table includes the values for the run that produced the median number of total draws.

\begin{table*}\centering
\ra{1.2}
\begin{tabular}{@{}rrrrcrrrcrrr@{}}\toprule
& \multicolumn{4}{c}{TAR curve \citep{Silk2013} } & \multicolumn{4}{c}{aABC-PMC}\\
\cmidrule{1-5} \cmidrule{6-10}
$t$ & $\epsilon_{t}$ & $\D_{t}$& $H\dist$ &&$t$ & $\epsilon_{t}$ & $q_{t}$  & $\D_{t}$& $H\dist$ \\\midrule
1 & 150 & 1,000 & 1.37   &&1 & 51.59 &  & 5,000 & 1.38 \\
2 & 51.26 & 11,560 & 1.26  &&2 & 51.02 & 0.19 & 8,130 & 1.36 \\
3 & $50.84$ & 1,403,040 & 0.174  &&3 & 51.00 & 0.16 & 99,596 & 0.68 \\
 &  &  &  &&4 & 39.33 & 0.17 & 138,972 & 0.43 \\
 &  &  &  &&5 & 0.07   & 0.06 & 32,045  & 0.067  \\
 &  &  &  &&6 & 0.00025  & 0.90  & 100,604  & 0.064 \\
\bottomrule
Total &  & 1,415,600 & &&  &  & & 384,347 \\
\end{tabular}
\caption{The number of draws needed in each iteration to reach $N=1,000$ accepted values for the ABC-PMC with the TAR curve-selected tolerances and the aABC-PMC algorithm.  (The displayed results were obtained by running the procedure 21 times and using the run that produced the median number of total draws.)  For the aABC-PMC algorithm, the quantile  automatically selected through the iterations is displayed under $q_{t}$. The procedure stopped once the quantile $q_7 = 0.9991$ was calculated. For the ABC--PMC algorithm a total of $1,415,600$ ($310$ sec.) draws are required, while our aABC-PMC takes $384,347$ ($258$ sec.) draws overall.
The number of draws listed for \cite{Silk2013} does \emph{not} include the draws required to build the TAR curve; however, we did include the TAR curve construction in the computational time.
}
\label{tab:Silk}
\end{table*}

It is apparent from Figure~\ref{fig:LocalMinEx} that the third iteration was an important step in which the large reduction of the tolerance allowed the algorithm to consider those few particles coming from the global optimal value at $\theta = 3$.
Although the raw tolerance hardly decreases between the first and the second iteration ($\epsilon_{1}=51.59$ and $\epsilon_{2}=51.02$), 
there is a substantial change between the ABC posteriors, from $\hat{\pi}_{\epsilon_{2}}$ to $\hat{\pi}_{\epsilon_{3}}$.
The majority of the accepted values from $t = 2$ are sampled near the local mode at $\theta = 10$, but the reduction resulting from the slightly smaller $\epsilon_3$ leads to the majority of values proposed near $\theta = 3$ to be accepted. 

\begin{figure}[htbp]
\centering
\subfloat[][{ABC posterior distributions \citep{Silk2013}}.]
{\includegraphics[width=2in]{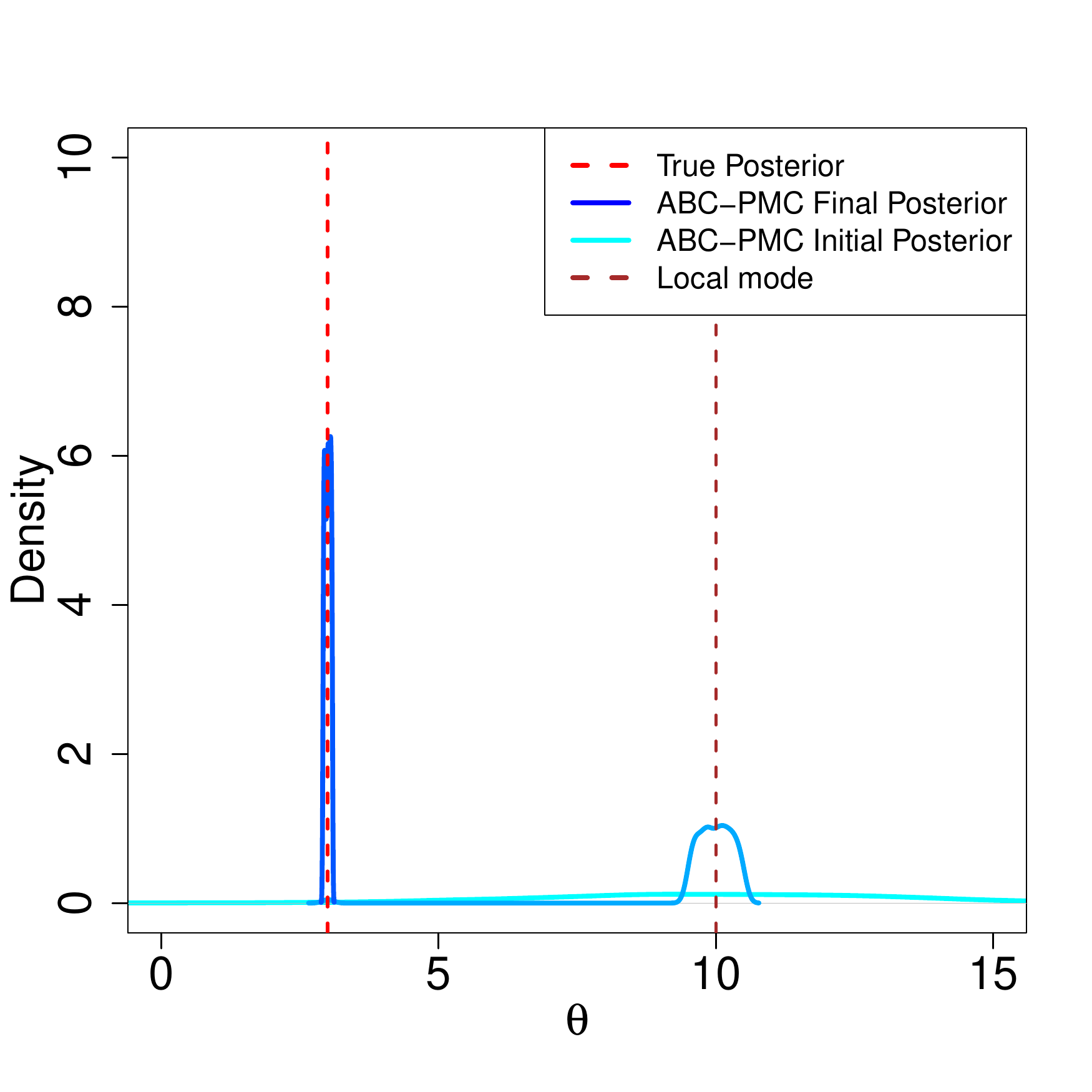}
\label{fig:ABCposteriorSilk}} \quad
\subfloat[][{aABC-PMC  posterior distributions}.]
{\includegraphics[width=2in]{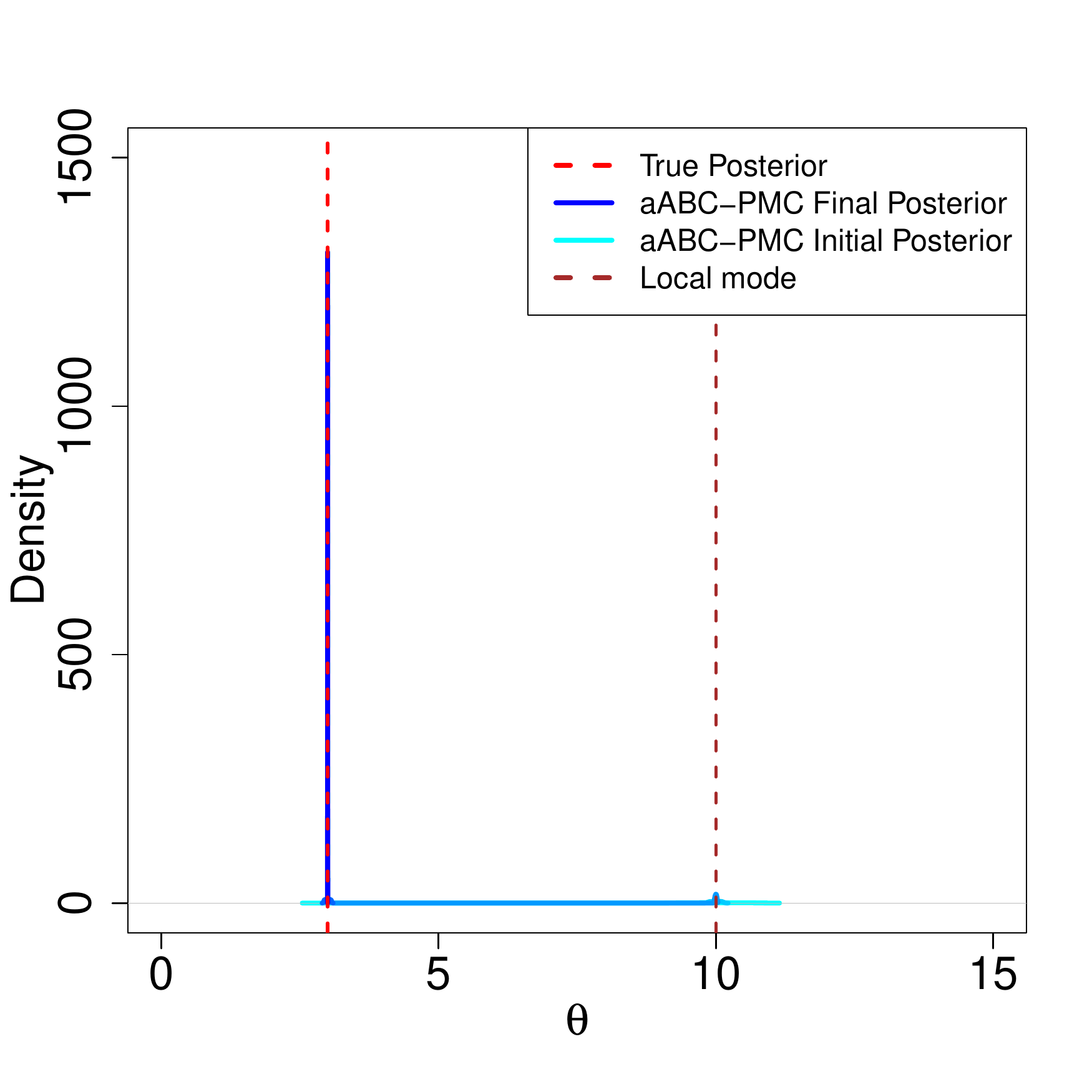}
\label{fig:aABCposteriorSilk}} \quad
\caption{ABC posterior distributions by iteration using (a) the TAR curve, and (b) the proposed aABC--PMC algorithm. The true posterior distribution, which is a Dirac function centered at $\theta=3$ is better captured by the aABC--PMC algorithm ($H\dist=0.064$), compared to the ABC--PMC method based on the TAR curve ($H\dist=0.17$). 
Note that the vertical axes are on different scales.}
\label{fig:SilkFinalPost}
\end{figure}

In order to compare the proposed aABC-PMC algorithm with the ABC-PMC approach of \cite{Silk2013} (see Section~\ref{sec:toleranceAndstop}), we estimated the TAR curve and the corresponding thresholds \citep{Silk2013}. The TAR curve is obtained by plotting on the $x$--axis several thresholds $\epsilon$ that might be picked for the next iteration of ABC simulations and on the $y$--axis their corresponding acceptance rates. The threshold $\epsilon$ recommended for the next ABC-PMC iteration is then selected by locating the ``elbow'' of the estimated TAR curve \citep{Silk2013}.
Since the forward model is computationally cheap, an approximation to the forward model was not needed. Instead, the TAR curve was estimated at each iteration by setting arbitrary grid points of tolerances having range in $(0, \epsilon_{t-1})$, 
running the ABC-PMC algorithm (for $t > 1$ the previous iteration's particle system and the Gaussian perturbation kernel are used),
and then calculating the acceptance rate according to Eq.\eqref{eq:real_eff}. This procedure was repeated 100 times and the resulting average TAR curve was used to retrieve the tolerance for the coming iteration, as was done in Fig. 2(left) of \cite{Silk2013}.
As result, a plot of acceptance rate vs. tolerances was obtained; the tolerance is set at the value corresponding to the elbow of the TAR curve. 
The series of tolerances, displayed together with the number of draws in Table \ref{tab:Silk}, is $\epsilon_{1:3}= (150, 51.26, 50.84)$ and the corresponding ABC posterior distributions are displayed in Figure \ref{fig:ABCposteriorSilk}.
The number of draws listed for \cite{Silk2013} does \emph{not} include the draws required to build the TAR curve; however, we did include the TAR curve construction in displayed computational time.
We note that the true posterior distribution, which is a Dirac function centered in $\theta=3$, is not suitably approximated by \cite{Silk2013} ($H\dist=0.17$). 

In order to calculate the Hellinger distance in this example, we approximate the true posterior (i.e., a Dirac function at $\theta=3$) with an $N$-dimensional vector with all elements equal to $3$.  

For $t=4$, the estimated TAR curve did not have an elbow and, consequently, there was no additional shrinkage of the tolerance resulting in an ABC posterior that was not a suitable approximation to the true posterior distribution; the final tolerance $\epsilon_3$ was too high.
We tried making adjustments to the TAR curve grid to see if this could be improved.  When using fewer grid points (e.g. 10) for the TAR curve, we were able to improve the performance.  However, this improved performance was due to poorer approximation to the TAR curve.  In general, it would be preferable if a better estimate of the TAR curve lead to better performance. A higher resolution TAR curve grid with 1000 grid points also was not able to find the global optimal solution.
In contrast, as shown in Figure \ref{fig:aABCposteriorSilk}, the proposed aABC-PMC approach provides a better approximation of the true posterior distribution although the number of draws required by the simulator is only of $384,347$ (compared to 1,415,600 draws required by ABC--PMC with the 100 point TAR curve grid).
%


\cite{Silk2013} note that if the particles are sampled from a large region of the parameter space that has a negligible mass in the posterior distribution, there is a risk of getting stuck in this parameter region if the tolerance is not selected carefully. In other words, the parameter space needs to be sufficiently explored in order to get enough particles in regions near the global optimal value.  
In the first iteration of the aABC-PMC algorithm the number of particles sampled directly from the prior was $kN$ with $k = 5$, which seems to work well in the examples considered. We emphasize that moving toward relevant regions of the parameter space needs to happen in the first few iterations of the ABC-PMC procedure, since uniformly small reductions in the tolerance sequence (e.g. using a fixed $q_t \ge 0.25$)  could end up removing those few important particles near the global optimal value, even if the number of particles sampled directly from the prior is $5N$.

The initial exploration of the parameter space and the definition of small enough quantiles in the first iterations appears to be why in the procedure based on the TAR curve, the total number of draws needed by the ABC--PMC algorithm is large, making it very expensive computationally.  In fact, at the end of the second iteration, the majority of the previous iteration's accepted particles are drawn near the local minimum. Moreover, since their $N\init=N$, only few candidates close to the global optimum are available. This means that when a particle is resampled, it will likely come from regions near to the local minimum and therefore it may be easily rejected during the third iteration of the ABC--PMC algorithm, for which the selected tolerance is $\epsilon_3=50.84$.

The proposed aABC-PMC algorithm allows for small $q_t$'s early on, when larger improvements occur between the sequential ABC posteriors. By doing so, larger reductions in the tolerance sequence can be taken in the first iterations of the ABC-PMC, which results in moving away from local optimal values into better regions of the parameter space.
If a sufficient reduction of the tolerance is not made early on, achieving a good approximation of the true posterior distribution is unlikely because the distances associated with the local optimal values will overwhelm the particle system so that it gets stuck in the local region.

As previously done for the Gaussian Mixture Model example presented in Sec.~\ref{sec:gmm}, we conclude the analysis of this model by performing a comparison between our adaptive aABC--PMC approach and the ABC-SMC algorithm by \cite{DelMoralEtAl2012}. Again, four initial particle sample sizes of $N\init$ are considered (N, 2N, 5N, and 10N) and 21 independent runs with the same dataset are performed. The results include the runs that produced the median number of total draws and are compared to the corresponding results obtained by our adaptive approach. The 5 parameters required by the ABC--SMC algorithm have been fixed as follows: $N=1,000$, $\epsilon=0.00025$, $\alpha=0.5$, $M=1$ and $\text{nb}_{\text{threshold}}=N/2$. We note again that default values are used for the last three parameters, following the suggestions by \cite{DelMoralEtAl2012}. The results of the analysis are summarized in Table \ref{tab:ABCSMC_Silk_Efficiency_Ninit} and the corresponding posterior distributions are displayed in Fig. \ref{fig:SilkEx_ABCSMC}. From Table \ref{tab:ABCSMC_Silk_Efficiency_Ninit} with $k=5$, although the number of total draws of the ABC--SMC algorithm is smaller than the corresponding total number of draws obtained by the adaptive aABC-PMC, our procedure is faster in terms of computational time. Moreover, the final ABC posterior distribution obtained by the aABC-PMC algorithm ($H\dist = 0.064$) matches the true posterior distribution better than the one obtained by the ABC-SMC sampler ($H\dist=0.388$). 
On the other hand, the ABC-SMC sampler successfully explores relevant regions of the parameter space for $k=1$ and $k=2$, while our aABC-PMC failed to reach the global mode for $k=1, 2$ because too few particles from the global mode were drawn in the first iteration of the procedure. However, the final ABC posterior distribution obtained by the aABC-PMC algorithm with the recommended $k=5$, and for which $H\dist = 0.064$, better matches the true posterior compared to any ABC posterior distribution obtained by the ABC-SMC algorithm (Figure \ref{fig:SilkEx_ABCSMC}).

\begin{table}
\begin{minipage}[c]{0.50\linewidth}
\centering
\resizebox{\columnwidth}{!}{%
\scalebox{1}{%
\begin{tabular}{@{}rrrrcrrrcrrr@{}}\toprule
\cmidrule{1-5}
 &  $\text{D}_{t}$ & time (sec)  & $H\dist$ &  $\epsilon_{t}$\\
N  & 56,535 & 31 & 0.374 & 0.00027  \\
2N  & 111,478 &  106 & 0.368 & 0.00050 \\
5N &  278,412 & 606  & 0.388 & 0.00017 \\
10N  & 521,945  & 2305 & 0.41 & 0.00022 \\
\bottomrule
\end{tabular}}
}
\caption{ABC-SMC algorithm with different choices for $N\init$ $(N, 2N, 5N, 10N)$ for the \cite{Silk2013}'s model. 
}
\label{tab:ABCSMC_Silk_Efficiency_Ninit}
\end{minipage}\hfill
\begin{minipage}[c]{0.48\linewidth}
\centering
\includegraphics[width=.95\columnwidth]{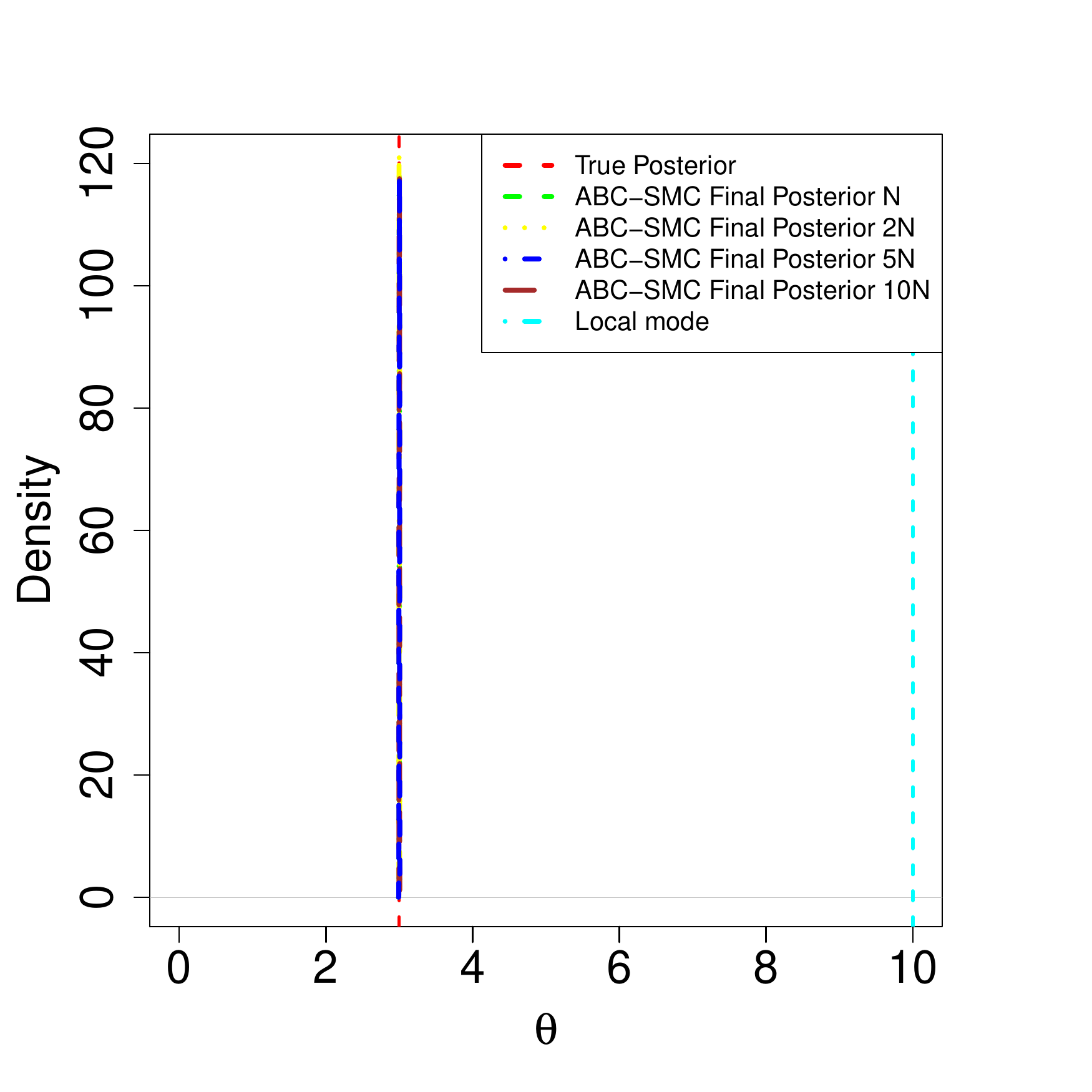}
\captionof{figure}{\hspace{-.1cm}: ABC-SMC final posterior distributions with different choices for $N\init$ $(N, 2N, 5N, 10N)$ for the \cite{Silk2013}'s model. 
}
\label{fig:SilkEx_ABCSMC}
\end{minipage}
\end{table}
\subsection{Bacterial Infection in Day Care Centers Example } \label{sec:daycarel}

The final model we consider, discussed by \cite{numminen2013estimating}, uses data on colonizations of the bacterium \textit{Streptococcus pneumoniae}. Discussion about mathematical models for such scenarios, known as household models, can be found in \cite{hoti2009outbreaks} or \cite{brooks2011epidemiologic}. According to the specifications provided in \cite{numminen2013estimating}, the transmission process is modeled with four parameters. Two parameters, $\beta$ and $\Lambda$, account for the hazards of infection from the day care center and from the community, respectively. Another parameter, $\theta$, scales the probability of co-infection.  Finally, the parameter $\gamma$ corresponds to the rate of clearance of an infection. In the following analyses we considered $\gamma=1$ fixed and known, to be consistent with the analysis in \cite{numminen2013estimating}.

The observed data consists of the identified pneumococcal strains in a total of 611 children from 29 day care centers, with varying numbers of sampled attendees per day \citep{vestrheim2008effectiveness, vestrheim2010impact}.  For each of the 29 day care centers, a binary matrix with varying number of sampled attendees is available. 
For each sampled attendee, the state of carrying one of the 33 different pneumococcal strains or not is indicated by a 1 or 0, respectively, in the binary matrix.
As pointed out in \cite{gutmann2016bayesian} statistical inference is challenging in this setting since the data represent a snapshot of the state of the sampled attendees at a single time point only. Moreover, the modeled system involves infinitely many correlated unobserved variables, since the modeled process evolves in continuous time. Using the observed colonizations with bacterial strains, the following four summary statistics are obtained for each of the 29 day care centers: the Shannon index of diversity of the distribution of the observed strains, the number of different strains, the prevalence of carriage among the observed individuals, and the prevalence of multiple infections among the observed individuals.  By doing so, the dimensionality of the problems reduces from a $611 \cdot 33 \cdot 29 = 584,727$ dimensional space to a $4 \cdot 29 =116$ dimensional space.

\cite{numminen2013estimating} use the four summary statistics and four tolerances, $\epsilon = (\epsilon_1, \epsilon_2, \epsilon_3, \epsilon_4)$, for each iteration of their procedure.
Instead, we use the approach of \cite{gutmann2016bayesian}. 
Each of the four summary statistics is rescaled so that the maximum value for each of the four the summary statistics is one. 
Then the summary statistics are vectorized in order to obtain a single vector of dimension $116$. 
Finally the $L^1$ distance between the vector corresponding to $y\pro$ and the vector corresponding to $y\obs$ is calculated, with the result divided by $116$. By doing so, only one tolerance is used in the ABC procedure.

The series of tolerances used in \cite{numminen2013estimating} was based on the ABC--Sequential Monte Carlo (ABC--SMC) method proposed by \cite{DelMoralEtAl2012}. The ABC--SMC method of \cite{DelMoralEtAl2012} adaptively proposes a series of tolerances by estimating, at the end of each iteration, the effective sample size (ESS). For a generic iteration $t$ the ESS is defined as:
\begin{equation}
\text{ESS}(\{W_t^{(J)}\}_{J=1}^{N})=\left(\sum_{J=1}^{N}\left(W_t^{(J)}\right)^2\right)^{-1},
\label{eq:ESS}
\end{equation}
where $W_t^{(J)}$ is the importance weight for particle $J = 1, \dots, N$ at iteration $t$ as defined in Eq.~\eqref{eq:importance_weights}. 
Once the ESS is estimated by using Eq.~\eqref{eq:ESS}, the new tolerance $\epsilon_{t+1}$ is obtained by solving the following for $\epsilon_{t+1}$:
 \begin{equation}
\text{ESS}(\{W_t^{(J)}\}_{J=1}^{N}, \epsilon_{t+1})= q_t \text{ESS}(\{W_{t-1}^{(J)}\}_{J=1}^{N}, \epsilon_{t}),
\label{eq:ESSepsilon}
\end{equation}
where $q_t$ is some pre-selected quantile which varies between 0 and 1. 
\cite{numminen2013estimating} had to adjust this to work for a their setting with four tolerances.
We note that our aABC--PMC approach does not require the specification of a quantile $q_t$, nor other parameters such as the number $M$ of simulations performed for each particle, the minimal effective sample size threshold below which a resampling of particles is performed, $\text{nb}_{\text{threshold}}$, and the final tolerance level,  $\epsilon_{\text{final}}$. Further details on the ABC--SMC algorithm and discussions on how to properly select its required parameters can be found in \cite{DelMoralEtAl2012}.


The prior distributions for the three parameters of interest are $\beta \sim \text{Unif}(0,11)$, $\Lambda \sim \text{Unif}(0,2)$, and $\theta \sim \text{Unif}(0,1)$. Starting from the second iteration of the ABC--PMC algorithm, proposals are perturbed with Gaussian kernels, using the specifications of \citet{BeaumontEtAl2009}. The desired particle sample size was set at $N=10,000$. For the aABC--PMC algorithm, the initial number of draws sampled from the prior distributions is set to $N\init = 5\times10,000$, in order to appropriately explore the parameter space.

The results of the analysis are summarized in Table \ref{tab:Numminen}, where the proposed adaptive rule for selecting the quantile performs better than the ABC--SMC algorithm both in terms of the computational time (3 days and 5 hours vs. 4 days and 12 hours using a cluster computer) and the total number of draws ($1,085,696$ draws vs.  $2,199,760$ draws). Because the proposed sampling procedure stops after $t=4$ iterations, the expensive forward model is used fewer times, achieving final posterior distributions in a shorter amount of time. 
We note that the number of particles sampled in the first iteration has an important role in the performance of the algorithm. In fact, having sampled from the priors $\D_{1}=50,000$ particles allowed the aABC--PMC algorithm to initiate with a smaller tolerance $\epsilon_1=1.26$ compared to the ABC--SMC algorithm ($\epsilon_1=3.91$ by fixing $\D_{1}=10,000$ particles).

\begin{table*}\centering
\ra{1.2}
\begin{tabular}{@{}rrrrcrrrcrrr@{}}\toprule
& \multicolumn{3}{c}{Numminen et al. (2013)} & \multicolumn{4}{c}{aABC-PMC}\\
\cmidrule{1-4} \cmidrule{5-8}
$t$ & $\epsilon_{t}$ & $\D_{t}$ &&$t$ & $\epsilon_{t}$ &  $q_{t}$ & $\D_{t}$ \\\midrule
1 & 3.91  & 10,000  && 1 & 1.26 &  & 50,000 \\
2 & 1.94  & 121,374  && 2 & 1.04  & 0.19 & 154,142  \\
3 & 1.28  & 277,997 && 3 & 0.97  & 0.31  & 489,239  \\
4 & 0.99 & 572,007  && 4 & 0.93  & 0.74 & 792,315 \\
5 & 0.84 & 1,218,760  &&  &   &  &  \\
\bottomrule
Total &  & 2,199,760 & &&  &  & 1,085,696 \\
\end{tabular}
\caption{Bacterial infection in day care centers results.  The number of draws needed in each iteration to reach $N=10,000$ accepted values for the ABC-SMC as presented in \cite{gutmann2016bayesian} and the proposed aABC-PMC algorithm. In the aABC-PMC algorithm also the quantile automatically selected through the iterations is available. The procedure stopped once the quantile $q_5 = 0.993$ was calculated. For the ABC--SMC algorithm a total of $2,199,760$ (4 days and 12 hours on a cluster with 200 cores) draws are required, while our aABC-PMC takes $1,085,696$ draws (3 days and 5 hours on a cluster with 200 cores).}
\label{tab:Numminen}
\end{table*}

\begin{figure}[htbp]
\centering
\subfloat[][{Final ABC posteriors for $\beta$}.]
{\includegraphics[height=.31\columnwidth]{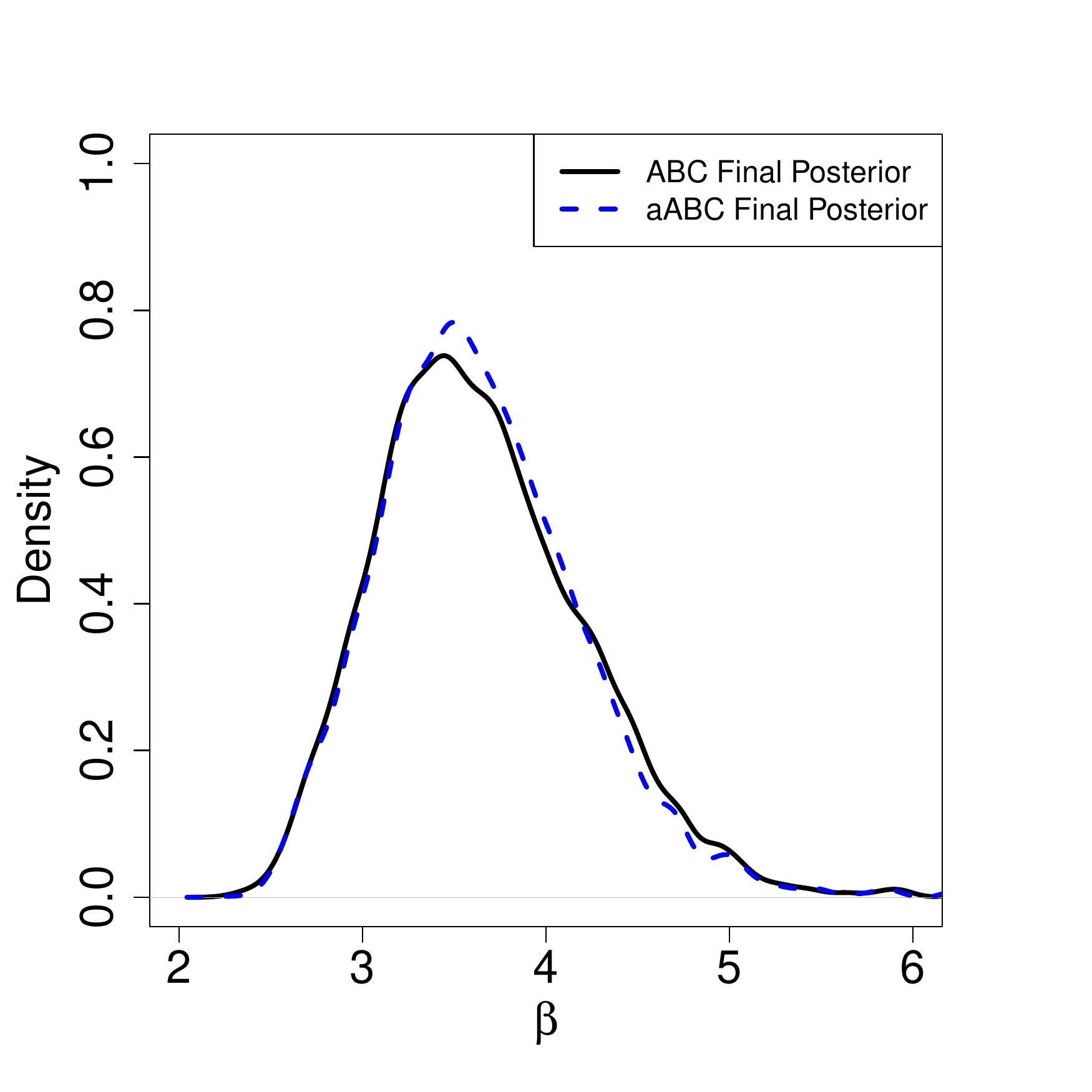}} \quad
\subfloat[][{Final ABC posteriors for $\lambda$}.]
{\includegraphics[height=.31\columnwidth]{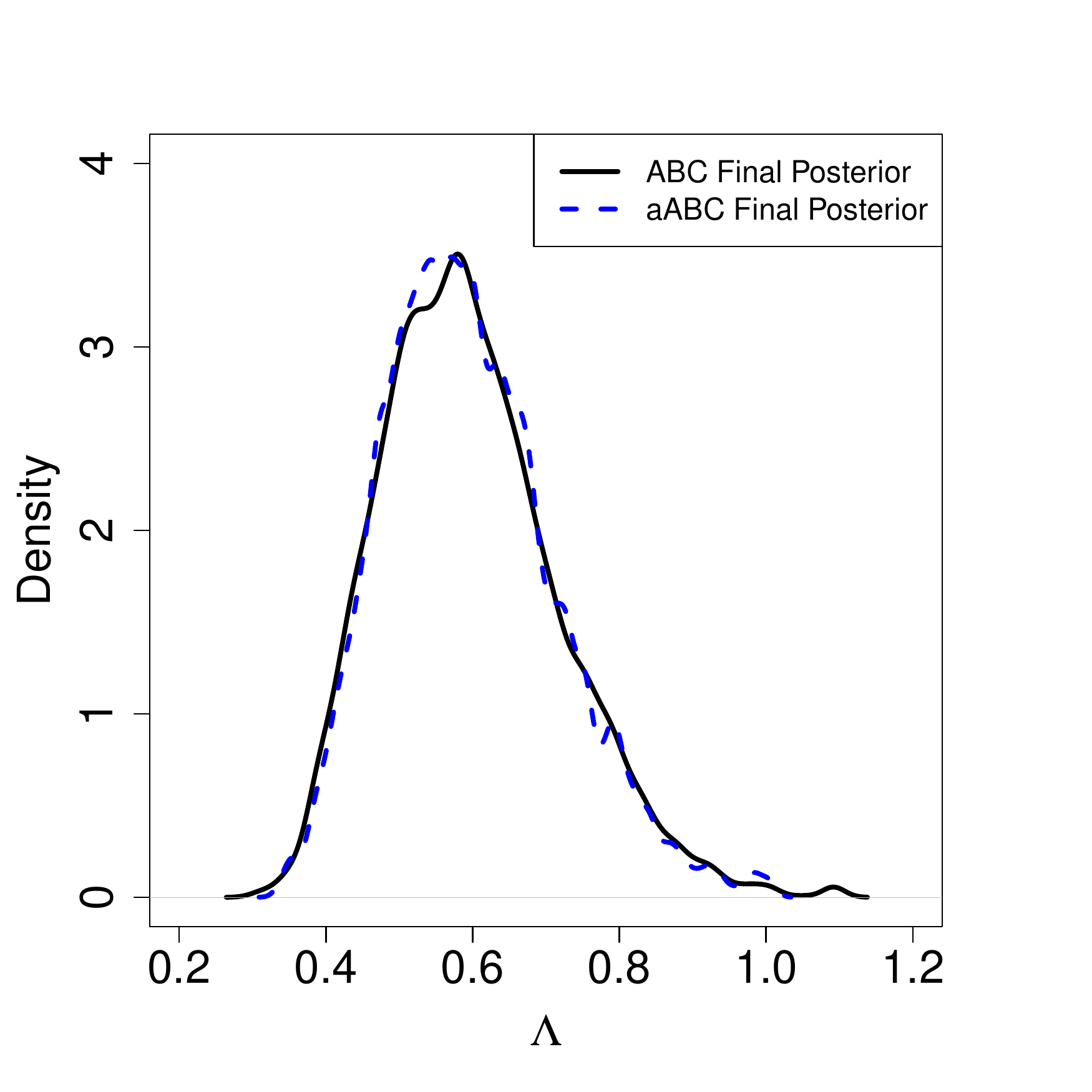}} \quad
\subfloat[][{Final ABC posteriors for $\theta$}.]
{\includegraphics[height=.31\columnwidth]{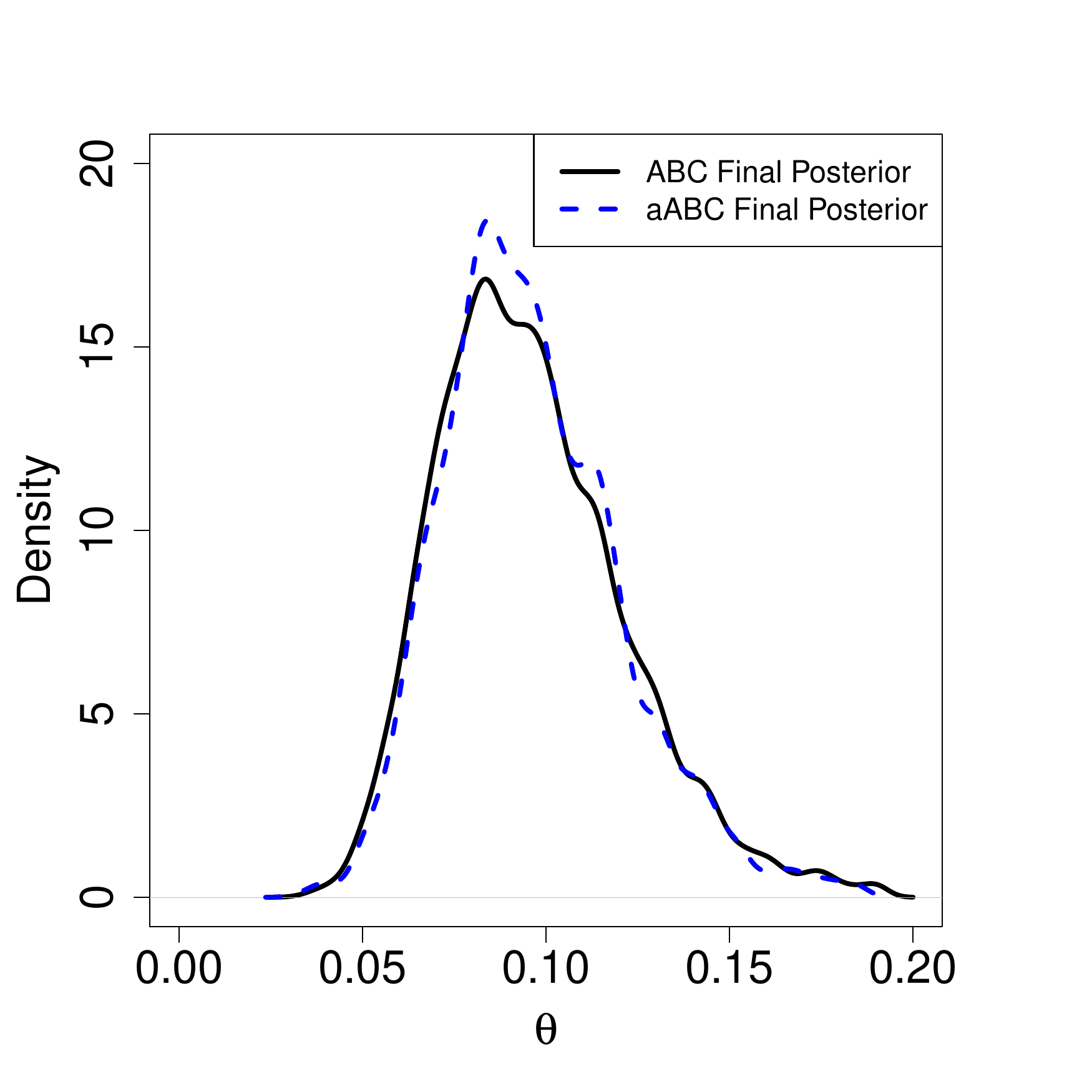}}
\caption{Bacterial infection in day care centers ABC posteriors. Comparison between the final posterior distributions for $\beta$, $\lambda$ and $\theta$ obtained by using \cite{DelMoralEtAl2012}'s adaptive selection of the tolerances (solid black) and by using the aABC-PMC algorithm (dashed blue). 
}
\label{fig:DCC}
\end{figure}

The ABC posteriors for the three parameters $\beta$, $\Lambda$ and $\theta$ for the tolerances of \cite{numminen2013estimating} selected by using ABC--SMC and the proposed aABC-PMC approach are displayed in Figures \ref{fig:DCC}.  We note that the final tolerance from \cite{numminen2013estimating}, $\epsilon_{5}=0.83$, is slightly smaller than the final tolerance of aABC-PMC, $\epsilon_{4}=0.93$, but the posteriors for $\beta$, $\Lambda$ and $\theta$ are comparable, with the Hellinger distances respectively equals to $H_{\text{dist}}=0.079, 0.097, 0.093$\footnote{The Hellinger distances are calculated between the ABC posterior distributions found by \cite{numminen2013estimating} and the corresponding ABC posterior distributions retrieved with our aABC-PMC approach.}.

\section{Concluding remarks} \label{sec:conclude}

The ABC-PMC algorithm of \citet{BeaumontEtAl2009} has lead to great improvements over the basic ABC rejection algorithm in terms of sampling efficiency.  
However, to use ABC-PMC it is necessary to define a sequence of tolerances along with the total number of iterations.  
We propose an approach leveraging ratio estimating methods for shrinking the tolerances by adaptively selecting a suitable quantile based on the progression of the estimated ABC posteriors. 
The proposed adjustment to the existing algorithm is shown to be able to deal with the possible presence of local modes and shrinks the tolerance in such a way that fewer draws are needed from the forward model compared to commonly used techniques for selecting the tolerances.
A simple criterion for stopping the algorithm based on the behavior of the sequential ABC posterior distribution is also presented.
The empirical performance in the examples considered suggests the proposed aABC-PMC algorithm is superior to the other options considered in terms of computational time and the number of draws from the forward model.  
Based on the computational experiments we envisage that the proposed aABC-PMC algorithm performs generally well when dealing with small to moderate dimensional problems for which the original ABC-PMC algorithm was developed. It remains as a challenge for the future research to generalize these samplers to higher dimensional models.

{\section{Supplementary Material}
Supplementary material for ``Adaptive Approximate Bayesian Computation Tolerance Selection'' \\ (DOI: ; .pdf).
}
\section{Acknowledgements}

The authors thank IT-University of Helsinki and Yale's Center for Research Computing for the computational resources provided to execute the analyses of the present work.
U. Simola was partially supported by Fondazione CARIPARO and supported by the Academy of Finland grant no. 1313197. J. Corander was supported by the ERC grant no. 742158. The authors are grateful for the comments and feedback from the anonymous associate editor and referees, which significantly helped to improve this work.
\bibliographystyle{ba}
\bibliography{mybib.bib}

\begin{thebibliography}{6}
\newcommand{\enquote}[1]{``#1''}
\expandafter\ifx\csname natexlab\endcsname\relax\def\natexlab#1{#1}\fi
\expandafter\ifx\csname url\endcsname\relax
  \def\url#1{{\tt #1}}\fi
\expandafter\ifx\csname urlprefix\endcsname\relax\def\urlprefix{URL }\fi
\ifx\endbibitem\undefined \let\endbibitem\relax\fi

\bibitem[{Del~Moral et~al.(2012)Del~Moral, Doucet, and
  Jasra}]{DelMoralEtAl2012}
Del~Moral, P., Doucet, A., and Jasra, A. (2012).
\newblock \enquote{An adaptive sequential Monte Carlo method for approximate
  Bayesian computation.}
\newblock {\em Statistics and Computing\/}, 22(5): 1009--1020.
\endbibitem

\bibitem[{J{\"a}rvenp{\"a}{\"a} et~al.(2016)J{\"a}rvenp{\"a}{\"a}, Gutmann,
  Vehtari, and Marttinen}]{jarvenpaa2016gaussian}
J{\"a}rvenp{\"a}{\"a}, M., Gutmann, M., Vehtari, A., and Marttinen, P. (2016).
\newblock \enquote{Gaussian process modeling in approximate Bayesian
  computation to estimate horizontal gene transfer in bacteria.}
\newblock {\em arXiv preprint arXiv:1610.06462\/}.
\endbibitem

\bibitem[{Lotka(1925)}]{Lotka}
Lotka, A. (1925).
\newblock {\em Elements of physical biology\/}, volume~1.
\newblock Baltimore, MD: Williams \& Wilkins Co.
\endbibitem

\bibitem[{Sedki et~al.(2012)Sedki, Pudlo, Marin, Robert, and
  Cornuet}]{sedki2012efficient}
Sedki, M., Pudlo, P., Marin, J.-M., Robert, C.~P., and Cornuet, J.-M. (2012).
\newblock \enquote{Efficient learning in ABC algorithms.}
\newblock {\em arXiv preprint arXiv:1210.1388\/}.
\endbibitem

\bibitem[{Toni et~al.(2009)Toni, Welch, Strelkowa, Ipsen, and
  Stumpf}]{TonyEtAl}
Toni, T., Welch, D., Strelkowa, N., Ipsen, A., and Stumpf, M. P.~H. (2009).
\newblock \enquote{Approximate Bayesian computation scheme for parameter
  inference and model selection in dynamical systems.}
\newblock {\em Journal of the Royal Society, Interface / the Royal Society\/},
  6(31): 187--202.
\endbibitem

\bibitem[{Volterra(1927)}]{Volterra:1927nr}
Volterra, V. (1927).
\newblock {\em Variazioni e fluttuazioni del numero d'individui in specie
  animali conviventi\/}.
\newblock C. Ferrari.
\endbibitem

\end{thebibliography}


\begin{thebibliography}{55}
\newcommand{\enquote}[1]{``#1''}
\expandafter\ifx\csname natexlab\endcsname\relax\def\natexlab#1{#1}\fi
\expandafter\ifx\csname url\endcsname\relax
  \def\url#1{{\tt #1}}\fi
\expandafter\ifx\csname urlprefix\endcsname\relax\def\urlprefix{URL }\fi
\ifx\endbibitem\undefined \let\endbibitem\relax\fi

\bibitem[{Andrieu et~al.(2003)Andrieu, De~Freitas, Doucet, and
  Jordan}]{andrieu2003introduction}
Andrieu, C., De~Freitas, N., Doucet, A., and Jordan, M.~I. (2003).
\newblock \enquote{An introduction to MCMC for machine learning.}
\newblock {\em Machine learning\/}, 50(1-2): 5--43.
\endbibitem

\bibitem[{Beaumont(2010)}]{Beaumont2010}
Beaumont, M.~A. (2010).
\newblock \enquote{Approximate bayesian computation in evolution and ecology.}
\newblock {\em Annual review of ecology, evolution, and systematics 41\/}, 96:
  379 -- 406.
\endbibitem

\bibitem[{Beaumont et~al.(2009)Beaumont, Cornuet, Marin, and
  Robert}]{BeaumontEtAl2009}
Beaumont, M.~A., Cornuet, J.-M., Marin, J.-M., and Robert, C.~P. (2009).
\newblock \enquote{Adaptive approximate {B}ayesian computation.}
\newblock {\em Biometrika\/}, 96(4): 983 -- 990.
\endbibitem

\bibitem[{Bickel et~al.(2007)Bickel, Br{\"u}ckner, and
  Scheffer}]{bickel2007discriminative}
Bickel, S., Br{\"u}ckner, M., and Scheffer, T. (2007).
\newblock \enquote{Discriminative learning for differing training and test
  distributions.}
\newblock In {\em Proceedings of the 24th international conference on Machine
  learning\/}, 81--88. ACM.
\endbibitem

\bibitem[{Blum et~al.(2013)Blum, Nunes, Prangle, and Sisson}]{BlumEtAl2013}
Blum, M., Nunes, M., Prangle, D., and Sisson, S. (2013).
\newblock \enquote{A comparative review of dimension reduction methods in
  approximate {B}ayesian computation.}
\newblock {\em Statistical Science\/}, 28(2): 189 -- 208.
\endbibitem

\bibitem[{Blum(2010)}]{Blum2010}
Blum, M.~G. (2010).
\newblock \enquote{Approximate {B}ayesian Computation: A nonparametric
  perspective.}
\newblock {\em Journal of American Statistical Association\/}, 105(491): 1178
  -- 1187.
\endbibitem

\bibitem[{Bonassi and West(2015)}]{Bonassi2015}
Bonassi, F. and West, M. (2015).
\newblock \enquote{Sequential Monte Carlo with Adaptive Weights for Approximate
  Bayesian Computation.}
\newblock {\em Bayesian Analysis\/}, (10): 171--187.
\endbibitem

\bibitem[{Bregman(1967)}]{bregman1967relaxation}
Bregman, L.~M. (1967).
\newblock \enquote{The relaxation method of finding the common point of convex
  sets and its application to the solution of problems in convex programming.}
\newblock {\em USSR computational mathematics and mathematical physics\/},
  7(3): 200--217.
\endbibitem

\bibitem[{Brent(2013)}]{brent2013algorithms}
Brent, R.~P. (2013).
\newblock {\em Algorithms for minimization without derivatives\/}.
\newblock Courier Corporation.
\endbibitem

\bibitem[{Brooks-Pollock et~al.(2011)Brooks-Pollock, Becerra, Goldstein, Cohen,
  and Murray}]{brooks2011epidemiologic}
Brooks-Pollock, E., Becerra, M.~C., Goldstein, E., Cohen, T., and Murray, M.~B.
  (2011).
\newblock \enquote{Epidemiologic inference from the distribution of
  tuberculosis cases in households in Lima, Peru.}
\newblock {\em Journal of Infectious Diseases\/}, 203(11): 1582--1589.
\endbibitem

\bibitem[{Cameron and Pettitt(2012)}]{CameronPettitt2012}
Cameron, E. and Pettitt, A.~N. (2012).
\newblock \enquote{Approximate Bayesian Computation for Astronomical Model
  Analysis: A Case Study in Galaxy Demographics and Morphological
  Transformation at High Redshift.}
\newblock {\em Monthly Notices of the Royal Astronomical Society\/}, 425:
  44--65.
\endbibitem

\bibitem[{Cisewski-Kehe et~al.(2019)Cisewski-Kehe, Weller, Schafer
  et~al.}]{Cisewski-Kehe:2019aa}
Cisewski-Kehe, J., Weller, G., Schafer, C., et~al. (2019).
\newblock \enquote{A preferential attachment model for the stellar initial mass
  function.}
\newblock {\em Electronic Journal of Statistics\/}, 13(1): 1580--1607.
\endbibitem

\bibitem[{Corander et~al.(2017)Corander, Fraser, Gutmann, Arnold, Hanage,
  Bentley, Lipsitch, and Croucher}]{corander2017frequency}
Corander, J., Fraser, C., Gutmann, M.~U., Arnold, B., Hanage, W.~P., Bentley,
  S.~D., Lipsitch, M., and Croucher, N.~J. (2017).
\newblock \enquote{Frequency-dependent selection in vaccine-associated
  pneumococcal population dynamics.}
\newblock {\em Nature ecology \& evolution\/}, 1(12): 1950.
\endbibitem

\bibitem[{Cornuet et~al.(2008)Cornuet, Santos, Beaumont, Robert, Marin,
  Balding, Guillemaud, and Estoup}]{Cornuet2008}
Cornuet, J., Santos, F., Beaumont, M., Robert, C., Marin, J., Balding, D.,
  Guillemaud, T., and Estoup, A. (2008).
\newblock \enquote{Inferring population history with DIY ABC: a user-friendly
  approach to Approximate Bayesian Computation.}
\newblock {\em Bioinformatics\/}.
\endbibitem

\bibitem[{Csill\'{e}ry et~al.(2010)Csill\'{e}ry, Blum, Gaggiotti, and
  Fran\c{c}ois}]{CsilleryEtAl2010}
Csill\'{e}ry, K., Blum, M.~G., Gaggiotti, O.~E., and Fran\c{c}ois, O. (2010).
\newblock \enquote{Approximate {B}ayesian Computation ({ABC}) in practice.}
\newblock {\em Trends in ecology \& evolution\/}, 25(7): 410 -- 418.
\endbibitem

\bibitem[{Del~Moral et~al.(2012)Del~Moral, Doucet, and
  Jasra}]{DelMoralEtAl2012}
Del~Moral, P., Doucet, A., and Jasra, A. (2012).
\newblock \enquote{An adaptive sequential Monte Carlo method for approximate
  Bayesian computation.}
\newblock {\em Statistics and Computing\/}, 22(5): 1009--1020.
\endbibitem

\bibitem[{Drovandi and Pettitt(2011)}]{DrovandiEtAl2011}
Drovandi, C.~C. and Pettitt, A.~N. (2011).
\newblock \enquote{Estimation of parameters for macroparasite population
  evolution using approximate Bayesian computation. Biometrics.}
\newblock {\em Statistics and Computing\/}, 67(1): 225--233.
\endbibitem

\bibitem[{Fearnhead and Prangle(2012)}]{FearnheadPrangle2012}
Fearnhead, P. and Prangle, D. (2012).
\newblock \enquote{Constructing summary statistics for approximate {B}ayesian
  computation: semi-automatic approximate {B}ayesian computation.}
\newblock {\em Journal of the Royal Statistical Society Series B\/}, 74(3):
  419--474.
\endbibitem

\bibitem[{Gelman et~al.(2014)Gelman, Carln, Stern, Dunson, Vehtari, and
  Rubin}]{GelmanEtAl2014}
Gelman, A., Carln, J., Stern, H., Dunson, D., Vehtari, A., and Rubin, D.
  (2014).
\newblock {\em Bayesian Data Analysis\/}.
\newblock Chapman \& Hall.
\endbibitem

\bibitem[{Gretton et~al.(2009)Gretton, Smola, Huang, Schmittfull, Borgwardt,
  and Sch{\"o}lkopf}]{gretton2009covariate}
Gretton, A., Smola, A.~J., Huang, J., Schmittfull, M., Borgwardt, K.~M., and
  Sch{\"o}lkopf, B. (2009).
\newblock \enquote{Covariate shift by kernel mean matching.}
\endbibitem

\bibitem[{Gutmann and Corander(2016)}]{gutmann2016bayesian}
Gutmann, M.~U. and Corander, J. (2016).
\newblock \enquote{Bayesian optimization for likelihood-free inference of
  simulator-based statistical models.}
\newblock {\em The Journal of Machine Learning Research\/}, 17(1): 4256--4302.
\endbibitem

\bibitem[{Hesterberg(1988)}]{hesterberg1988advances}
Hesterberg, T.~C. (1988).
\newblock \enquote{Advances in importance sampling.}
\newblock Ph.D. thesis, Stanford University.
\endbibitem

\bibitem[{Hido et~al.(2011)Hido, Tsuboi, Kashima, Sugiyama, and
  Kanamori}]{hido2011statistical}
Hido, S., Tsuboi, Y., Kashima, H., Sugiyama, M., and Kanamori, T. (2011).
\newblock \enquote{Statistical outlier detection using direct density ratio
  estimation.}
\newblock {\em Knowledge and information systems\/}, 26(2): 309--336.
\endbibitem

\bibitem[{Hoti et~al.(2009)Hoti, Er{\"a}st{\"o}, Leino, and
  Auranen}]{hoti2009outbreaks}
Hoti, F., Er{\"a}st{\"o}, P., Leino, T., and Auranen, K. (2009).
\newblock \enquote{Outbreaks of Streptococcus pneumoniae carriage in day care
  cohorts in Finland--implications for elimination of transmission.}
\newblock {\em BMC infectious diseases\/}, 9(1): 102.
\endbibitem

\bibitem[{Ishida et~al.(2015)Ishida, Vitenti, Penna-Lima, Cisewski, de~Souza,
  Trindade, Cameron et~al.}]{IshidaEtAl2015}
Ishida, E., Vitenti, S., Penna-Lima, M., Cisewski, J., de~Souza, R., Trindade,
  A., Cameron, E., et~al. (2015).
\newblock \enquote{cosmoabc: Likelihood-free inference via Population Monte
  Carlo Approximate Bayesian Computation.}
\newblock {\em Astronomy \& Computing\/}, 13: 1--11.
\endbibitem

\bibitem[{J{\"a}rvenp{\"a}{\"a} et~al.(2016)J{\"a}rvenp{\"a}{\"a}, Gutmann,
  Vehtari, and Marttinen}]{jarvenpaa2016gaussian}
J{\"a}rvenp{\"a}{\"a}, M., Gutmann, M., Vehtari, A., and Marttinen, P. (2016).
\newblock \enquote{Gaussian process modeling in approximate Bayesian
  computation to estimate horizontal gene transfer in bacteria.}
\newblock {\em arXiv preprint arXiv:1610.06462\/}.
\endbibitem

\bibitem[{Jennings and Madigan(2016)}]{JenningsEtAl}
Jennings, E. and Madigan, M. (2016).
\newblock \enquote{astroABC : An Approximate Bayesian Computation Sequential
  Monte Carlo sampler for cosmological parameter estimation.}
\newblock {\em Astronomy and Computing\/}.
\endbibitem

\bibitem[{Jennings et~al.(2016)Jennings, Wolf, and Sako}]{JenningsEtAl2}
Jennings, E., Wolf, R., and Sako, M. (2016).
\newblock \enquote{A new approach for obtaining cosmological constraints from
  type IA supernovae using approximate bayesian computation.}
\newblock {\em Astronomy and Computing\/}.
\endbibitem

\bibitem[{Joyce and Marjoram(2008)}]{JoyceMarjoram2008}
Joyce, P. and Marjoram, P. (2008).
\newblock \enquote{Approximately sufficient statistics and {B}ayesian
  computation.}
\newblock {\em Statistical Applications in Genetics and Molecular Biology\/},
  7(1): 1 -- 16.
\endbibitem

\bibitem[{Julier et~al.(2000)Julier, Uhlmann, and
  Durrant-Whyte}]{Julier:2000fk}
Julier, S., Uhlmann, J., and Durrant-Whyte, H.~F. (2000).
\newblock \enquote{A new method for the nonlinear transformation of means and
  covariances in filters and estimators.}
\newblock {\em IEEE Transactions on automatic control\/}, 45(3): 477--482.
\endbibitem

\bibitem[{Lenormand et~al.(2013)Lenormand, Jabot, and Deuant}]{Lenormand2013}
Lenormand, M., Jabot, F., and Deuant, G. (2013).
\newblock \enquote{Adaptive approximate bayesian computation for complex
  models.}
\newblock {\em Computational Statistics\/}, 6(28): 2777--2796.
\endbibitem

\bibitem[{Lintusaari et~al.(2017)Lintusaari, Gutmann, Dutta, Kaski, and
  Corander}]{lintusaari2017fundamentals}
Lintusaari, J., Gutmann, M.~U., Dutta, R., Kaski, S., and Corander, J. (2017).
\newblock \enquote{Fundamentals and recent developments in approximate Bayesian
  computation.}
\newblock {\em Systematic biology\/}, 66(1): e66--e82.
\endbibitem

\bibitem[{Marin et~al.(2012)Marin, Pudlo, Robert, and Ryder}]{MarinEtAl2012}
Marin, J.-M., Pudlo, P., Robert, C.~P., and Ryder, R.~J. (2012).
\newblock \enquote{Approximate {B}ayesian computational methods.}
\newblock {\em Statistics and Computing\/}, 22(6): 1167 -- 1180.
\endbibitem

\bibitem[{McKinley et~al.(2009)McKinley, Cook, and Deardon}]{McKinleyEtAl2009}
McKinley, T., Cook, A., and Deardon, R. (2009).
\newblock \enquote{Inference in epidemic models without likelihoods.}
\newblock {\em The International Journal of Biostatistics\/}, 171(5).
\endbibitem

\bibitem[{Numminen et~al.(2013)Numminen, Cheng, Gyllenberg, and
  Corander}]{numminen2013estimating}
Numminen, E., Cheng, L., Gyllenberg, M., and Corander, J. (2013).
\newblock \enquote{Estimating the transmission dynamics of Streptococcus
  pneumoniae from strain prevalence data.}
\newblock {\em Biometrics\/}, 69(3): 748--757.
\endbibitem

\bibitem[{Pritchard et~al.(1999)Pritchard, Seielstad, and
  Perez-Lezaun}]{PitchardEtAl1999}
Pritchard, J.~K., Seielstad, M.~T., and Perez-Lezaun, A. (1999).
\newblock \enquote{Population Growth of Human {Y} Chromosomes: A study of {Y}
  Chromosome Microsatellites.}
\newblock {\em Molecular Biology and Evolution\/}, 16(12): 1791 -- 1798.
\endbibitem

\bibitem[{{R Core Team}(2019)}]{Rcite}
{R Core Team} (2019).
\newblock {\em R: A Language and Environment for Statistical Computing\/}.
\newblock R Foundation for Statistical Computing, Vienna, Austria.
\newline\urlprefix\url{https://www.R-project.org/}
\endbibitem

\bibitem[{Ratmann et~al.(2013)Ratmann, Camacho, Meijer, and
  Donker}]{RatmannEtAl2013}
Ratmann, O., Camacho, A., Meijer, A., and Donker, G. (2013).
\newblock \enquote{Statistical modelling of summary values leads to accurate
  Approximate {B}ayesian computations.}
\newblock Unpublished.
\endbibitem

\bibitem[{Robert and Casella(2013)}]{robert2013monte}
Robert, C. and Casella, G. (2013).
\newblock {\em Monte Carlo statistical methods\/}.
\newblock Springer Science \& Business Media.
\endbibitem

\bibitem[{Rubin(1984)}]{Rubin1984}
Rubin, D.~B. (1984).
\newblock \enquote{Bayesianly justifiable and relevant frequency calculations
  for the applied statistician.}
\newblock {\em The Annals of Statistics\/}, 12(4): 1151--1172.
\endbibitem

\bibitem[{Schafer and Freeman(2012)}]{ShaferFreeman2012}
Schafer, C.~M. and Freeman, P.~E. (2012).
\newblock {\em Statistical Challenges in Modern Astronomy {V}\/}, chapter~1, 3
  -- 19.
\newblock Lecture Notes in Statistics. Springer.
\endbibitem

\bibitem[{Silk et~al.(2013)Silk, Filippi, and Stumpf}]{Silk2013}
Silk, D., Filippi, S., and Stumpf, M. (2013).
\newblock \enquote{Optimizing threshold-schedules for sequential approximate
  Bayesian computation: applications to molecular systems.}
\newblock {\em Statistical Applications in Genetics and Molecular Biology\/},
  5(12): 603--618.
\endbibitem

\bibitem[{Silverman(1986)}]{silverman1986density}
Silverman, B.~W. (1986).
\newblock {\em Density estimation for statistics and data analysis\/},
  volume~26.
\newblock CRC press.
\endbibitem

\bibitem[{Silverman(2018)}]{silverman2018density}
--- (2018).
\newblock {\em Density estimation for statistics and data analysis\/}.
\newblock Routledge.
\endbibitem

\bibitem[{Simola et~al.(2019)Simola, Pelssers, Barge, Conrad, and
  Corander}]{simola2019machine}
Simola, U., Pelssers, B., Barge, D., Conrad, J., and Corander, J. (2019).
\newblock \enquote{Machine learning accelerated likelihood-free event
  reconstruction in dark matter direct detection.}
\newblock {\em Journal of Instrumentation\/}, 14(03): P03004.
\endbibitem

\bibitem[{Sisson et~al.(2007)Sisson, Fan, and Tanaka}]{SissonEtAl2007}
Sisson, S.~A., Fan, Y., and Tanaka, M.~M. (2007).
\newblock \enquote{Sequential {M}onte {C}arlo without likelihoods.}
\newblock {\em Proceedings of the National Academy of Science\/}, 104(6): 1760
  -- 1765.
\endbibitem

\bibitem[{Sugiyama et~al.(2008)Sugiyama, Nakajima, Kashima, Buenau, and
  Kawanabe}]{sugiyama2008direct}
Sugiyama, M., Nakajima, S., Kashima, H., Buenau, P.~V., and Kawanabe, M.
  (2008).
\newblock \enquote{Direct importance estimation with model selection and its
  application to covariate shift adaptation.}
\newblock In {\em Advances in neural information processing systems\/},
  1433--1440.
\endbibitem

\bibitem[{Sugiyama et~al.(2010)Sugiyama, Suzuki, and
  Kanamori}]{sugiyama2010density}
Sugiyama, M., Suzuki, T., and Kanamori, T. (2010).
\newblock \enquote{Density Ratio Estimation: A Comprehensive Review
  (Statistical Experiment and Its Related Topics).}
\endbibitem

\bibitem[{Sugiyama et~al.(2012)Sugiyama, Suzuki, and
  Kanamori}]{sugiyama2012density}
--- (2012).
\newblock {\em Density ratio estimation in machine learning\/}.
\newblock Cambridge University Press.
\endbibitem

\bibitem[{Tavar\'e et~al.(1997)Tavar\'e, Balding, Griffiths, and
  Donnelly}]{TavareEtAl1997}
Tavar\'e, S., Balding, D.~J., Griffiths, R., and Donnelly, P. (1997).
\newblock \enquote{Inferring coalescence times from {DNA} sequence data.}
\newblock {\em Genetics\/}, 145: 505 -- 518.
\endbibitem

\bibitem[{Thornton and Andolfatto(2006)}]{ThorntonEtAl2006}
Thornton, K. and Andolfatto, P. (2006).
\newblock \enquote{Inference in epidemic models without likelihoods.}
\newblock {\em Genetics\/}, 172: 1607 -- 1619.
\endbibitem

\bibitem[{Toni et~al.(2009)Toni, Welch, Strelkowa, Ipsen, and
  Stumpf}]{TonyEtAl}
Toni, T., Welch, D., Strelkowa, N., Ipsen, A., and Stumpf, M. P.~H. (2009).
\newblock \enquote{Approximate Bayesian computation scheme for parameter
  inference and model selection in dynamical systems.}
\newblock {\em Journal of the Royal Society, Interface / the Royal Society\/},
  6(31): 187--202.
\endbibitem

\bibitem[{Vestrheim et~al.(2010)Vestrheim, H{\o}iby, Aaberge, and
  Caugant}]{vestrheim2010impact}
Vestrheim, D.~F., H{\o}iby, E.~A., Aaberge, I.~S., and Caugant, D.~A. (2010).
\newblock \enquote{Impact of a pneumococcal conjugate vaccination program on
  carriage among children in Norway.}
\newblock {\em Clin. Vaccine Immunol.\/}, 17(3): 325--334.
\endbibitem

\bibitem[{Vestrheim et~al.(2008)Vestrheim, L{\o}voll, Aaberge, Caugant,
  H{\o}iby, Bakke, and Bergsaker}]{vestrheim2008effectiveness}
Vestrheim, D.~F., L{\o}voll, {\O}., Aaberge, I.~S., Caugant, D.~A., H{\o}iby,
  E.~A., Bakke, H., and Bergsaker, M.~R. (2008).
\newblock \enquote{Effectiveness of a 2+ 1 dose schedule pneumococcal conjugate
  vaccination programme on invasive pneumococcal disease among children in
  Norway.}
\newblock {\em Vaccine\/}, 26(26): 3277--3281.
\endbibitem

\bibitem[{Weyant et~al.(2013)Weyant, Schafer, and Wood-Vasey}]{WeyantEtAl2013}
Weyant, A., Schafer, C., and Wood-Vasey, W.~M. (2013).
\newblock \enquote{Likelihood-free cosmological inference with type {I}a
  supernovae: approximate {B}ayesian computation for a complete treatment of
  uncertainty.}
\newblock {\em The Astrophysical Journal\/}, 764: 116.
\endbibitem

\end{thebibliography}

\end{document}